\documentclass[reprint, amsmath,amssymb,aps,pra,]{revtex4-2}

\usepackage{graphicx}
\usepackage{dcolumn}
\usepackage{bm}
\usepackage{subfigure}
\usepackage{caption}
\usepackage[dvipsnames]{xcolor}
\usepackage{hyperref}

\begin {document}
\preprint{APS/123-QED}
\title{Kinetics Of Vapor-Liquid And Vapor-Solid Phase Separation Under Gravity}
\author{Daniya Davis}
\author{Bhaskar Sen Gupta}%
 \email{bhaskar.sengupta@vit.ac.in}
\affiliation{%
 Department of Physics, School of Advanced Sciences, Vellore Institute of Technology, Vellore, Tamil Nadu 632014, India
}%

\date{\today}

\begin{abstract}
We study the kinetics of vapor-liquid and vapor-solid phase separation of a hydrodynamics preserving three-dimensional one component Lennard Jones system in the presence of external gravitational field using extensive molecular dynamic simulation. A bicontinuous domain structure is formed when the homogeneous system near the critical density is quenched inside the coexistence region. In the absence of gravity, the domain morphology is statistically self-similar and the length scale grows as per the existing laws. However, the presence of gravity destroys the isotropy of the system and affects the scaling laws. We observe an accelerated domain growth in the direction of the field at late time which resembles sedimentation process. Consequently, a new length scale emerges which strongly depends on the field strength. Similar behavior is observed in the direction perpendicular to the applied field, with a different growth rate. Finally, the validity of Porod's law and Superuniversality in such anisotropic systems is verified in terms of two-point equal time order parameter correlation function and static structure factor. 
\end{abstract}

\maketitle

\section{\label{sec:level1}Introduction}

If a homogeneous system is cooled rapidly within the miscibility gap, it becomes thermodynamically unstable and gradually separates into particle-rich and particle-poor domains. The separated regions continue to grow over time until the system reaches a state of local equilibrium or saturation \cite{Binder,Siggia,Furukawa,Miguel,Tanaka,Beysens,Tanaka1,Kendon,Puri,
Dutt,Laradji,Thakre,Ahmad}. This process of phase separation has numerous applications in various fields such as material science, biochemistry, chemical engineering \cite{Fisher,Stanley,Binder1,Jones,Bray}, and the petroleum industry \cite{Chen,Worrell,Nomura}. It is involved in the separation of various chemicals, such as the distillation of crude oil, as well as in the extraction of oil, natural gases and other resources. Phase separation also plays a crucial role in regulating the morphology and mechanical characteristics of polymers during their production.

Gaining a comprehensive understanding of the non-equilibrium evolution that occurs during phase separation in a system holds significant importance and attracts substantial research interest, both in theoretical and experimental domains. 

It is now well accepted that the domain growth pattern or domain morphology is characterized by a solitary length parameter that varies with time, denoted as $\ell(t)$ \cite{Bray}. This length is obtained from the equal time correlation function $C(\vec r,t)$, where $\vec r$ is the distance between two spatial points and t is the time after quench. The average domain size of the system follows the power law $\ell(t)  \sim t^{\alpha} $, where $\alpha$ is the universal growth exponent \cite{Siggia}. The exponent value is determined by the corresponding coarsening mechanism that drives phase separation. 

The growth behavior observed in a specific system depends on various factors, including temperature, composition, interfacial tension, and molecular interactions. In the phase separation process involving solid mixtures, diffusion of particles dominates and the corresponding growth exponent $\alpha=1/3$, predicted by Lifshitz and Slyozov \cite{Bray,Binder2}. But in the liquid-liquid or vapor-liquid phase separation, hydrodynamic effect plays a  significant role and the growth exponent changes accordingly. In such systems the diffusive phase is very transient and we observe a quick transition to the viscous hydrodynamic regime with $\alpha =1$. Finally, at late times the system enters into inertial hydrodynamics where the exponent is $\alpha =2/3$ \cite{Siggia,Furukawa}.

Numerous studies dedicated their focus on the phase separation of vapor-liquid systems at both near critical as well as off-critical densities and the subject is now well understood \cite{Majumdar,Jiarul,Roy1,Roy2}.
Even though infrequently but occasionally studies have been conducted on the vapor-solid systems and the kinetics at various densities have been explored \cite{Midya,Saikat1}. However, the nonequilibrium phenomena of such systems in the presence of an external field such as gravity is still in its infancy. Gravitational effect becomes important for situations when heavier domains situated atop lighter ones become unstable, leading to an acceleration in domain growth. In both the systems, one of the phases is heavier than the other as differentiated by gravity. Here, we investigate the spinodal decomposition and the phase ordering in vapor-liquid and vapor-solid phase separation under the influence of gravitational force field. More specifically, our main objective is to study the effect of gravity on the domain growth laws in different regimes mentioned above. Finally, we study the final state of the system, where the two phases are completely separated with an interface between them.

Fewer attempts were made to understand the effect of gravity using theory and computer simulations. For example, numerical study on the modified Cahn-Hilliard equation constructed from the master equation approach was carried out in two-dimension ~\cite{Puri1,Puri2}. An accelerated domain growth was found in the direction of applied gravitational field, and the growth was linear in time. The thermal fluctuation and hydrodynamic effects were not taken into account in this work. In another study, the front and domain growth was studied from the time-dependent Ginzburg-Landau equation with gravity ~\cite{Lacasta}. The diffusion coefficient was chosen to be concentration dependent. The growth exponents were estimated using scaling arguments and was compared with the numerical results in two dimension. The coarsening dynamics of the binary liquid mixture with density mismatch between the domains was studied numerically by solving the coupled Cahn–Hilliard/Navier–Stokes system of equations in the presence of gravity ~\cite{Badalassi}. The outcome demonstrated that even a slight density mismatch can have a notable impact on scaling laws due to gravitational effects. The coarsening was faster in the direction of applied field, and multiple length scale emerged in this process. Recently, the vapor-liquid phase separation of a van der Waals fluid described by the Navier-Stokes and the continuity equation subjected to gravitational force was studied numerically by applying the lattice Boltzmann method in two dimension. The domain growth along the gravity direction was characterized by the exponent $\alpha=1$. Surprisingly, the growth exponent was found to be insensitive to the strength of gravitational field ~\cite{Cristea}.

To the best of our knowledge, there is no atomistic simulation work on the kinetics of vapor-liquid phase separation of systems exposed to gravitational field. On the other hand, the vapor-solid phase separation with gravity remains completely unexplored. In this paper, we undertake molecular dynamics simulation study of vapor-liquid and vapor-solid phase separation in a one component fluid in three dimension with gravity. We consider a more realistic model from the experimental prospective where hydrodynamics is naturally incorporated in the system. In the presence of gravitational field, the domain growth is expected to be anisotropic. The main focus of this paper is to examine in detail the anisotropic scaling laws at different time regimes under the influence of various external field strength.

\section{Models And Methods}
For the present study, we consider a single component system of volume $V$ in three-dimension, consisting of $N$ particles interacting via the Lennard Jones (LJ) interaction potential given by,
\begin{equation}
\label{eq:lj_potential}
U(r)=4\epsilon\left[\left(\frac{\sigma}{ r}\right)^{12}- \left(\frac{\sigma}{r}\right)^6\right]
\end{equation}
Here $\epsilon$ denotes the interaction strength, $\sigma$ is the particle diameter and $r=|\vec r_i-\vec r_j|$ is the scalar distance between the two particles $i$ and $j$. To improve computational efficiency the range of interaction is truncated at $r=r_c=2.5\sigma$. The discontinuity in the potential and the corresponding force due to the insertion of this cut off is resolved by modifying the potential as follows, 
\begin{equation}
\label{eq:mod_potential}
u(r)=U(r)-U(r_c)-(r-r_c)\left(\frac{dU}{dr}\right)|_{r=r_c} 
\end{equation}
The distance and temperature are measured in units of $\sigma$ and $\epsilon/k_B$ respectively, where $k_B$ is Boltzmann's constant. For computational convenience mass of each particle, $\epsilon$, $\sigma$, $k_B$ are set to unity. The critical temperature and critical density for this model exhibiting vapor-liquid transition are  $T_c=0.94$ and $\rho_c=0.32$ respectively \cite{Jiarul}.   

We consider an elongated box with dimensions $L_x \times L_y \times L_z = 72\sigma \times 72\sigma \times 144\sigma$ having reflecting walls at the top and bottom in the z direction. Periodic boundary conditions are applied along the horizontal directions. The simulation box consists of 223948 particles which correspond to the near critical density of $\rho=0.3$.

The coarsening dynamics of the system is studied using molecular dynamics simulation (MD) performed in the canonical ensemble ($NVT$). Since we are dealing with fluids, it is preferable to use a hydrodynamics preserving thermostat to keep the temperature at the desired value. Therefore, throughout the current study, we use Nose-Hoover Thermostat which is well known for temperature controlling and at the same time preserving the hydrodynamics of the system \cite{Nose}. Velocity-Verlet algorithm is used to integrate the equation of motion for each particle that evaluates the position and velocity during each time step of the MD simulation \cite{Verlet}. The time step is chosen as  $\Delta t=0.005$, where time is expressed in units of ${(m\sigma^2/\epsilon)}^{1/2}$. 

\begin{figure}[ht]
    \centering
    \begin{subfigure}
        \centering
        \includegraphics[width=0.45\linewidth]{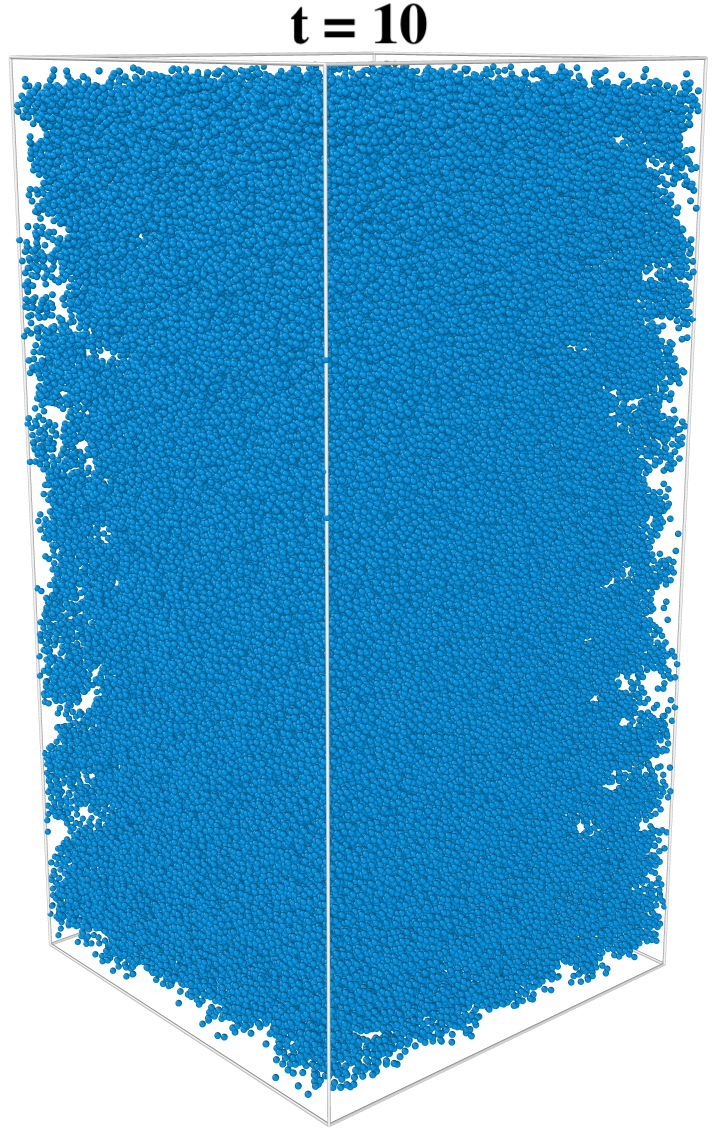}
     \end{subfigure}
    \begin{subfigure}
        \centering
        \includegraphics[width=0.45\linewidth]{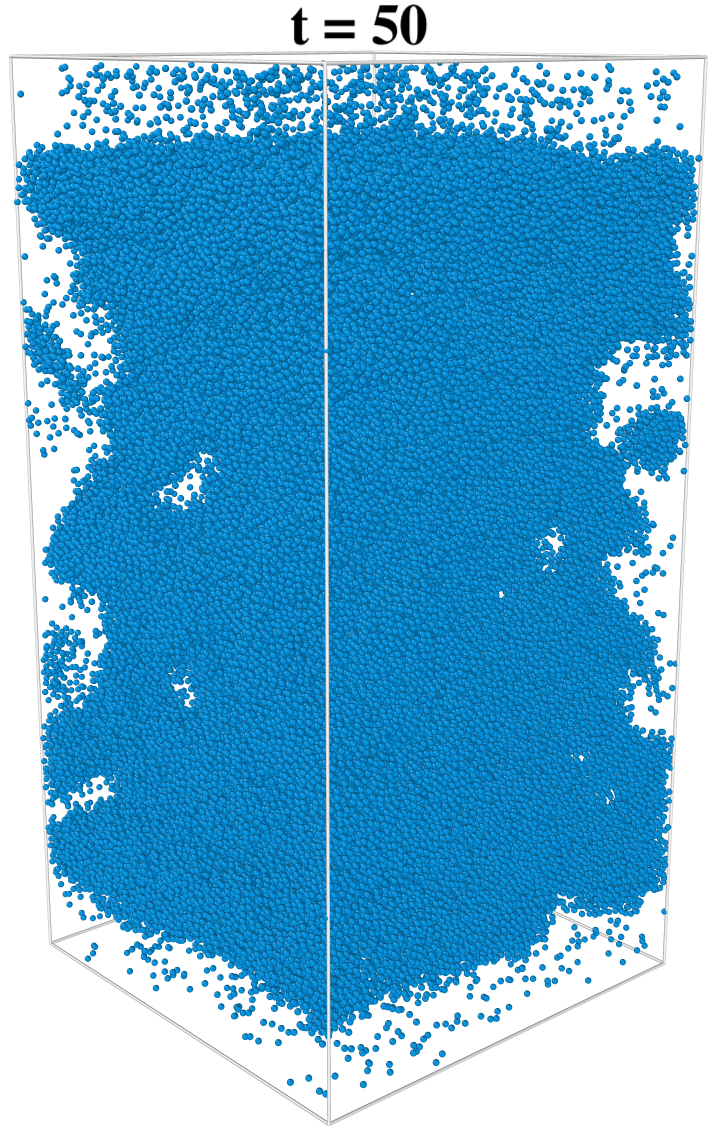}
    \end{subfigure}\\
    \begin{subfigure}
        \centering        
        \includegraphics[width=0.45\linewidth]{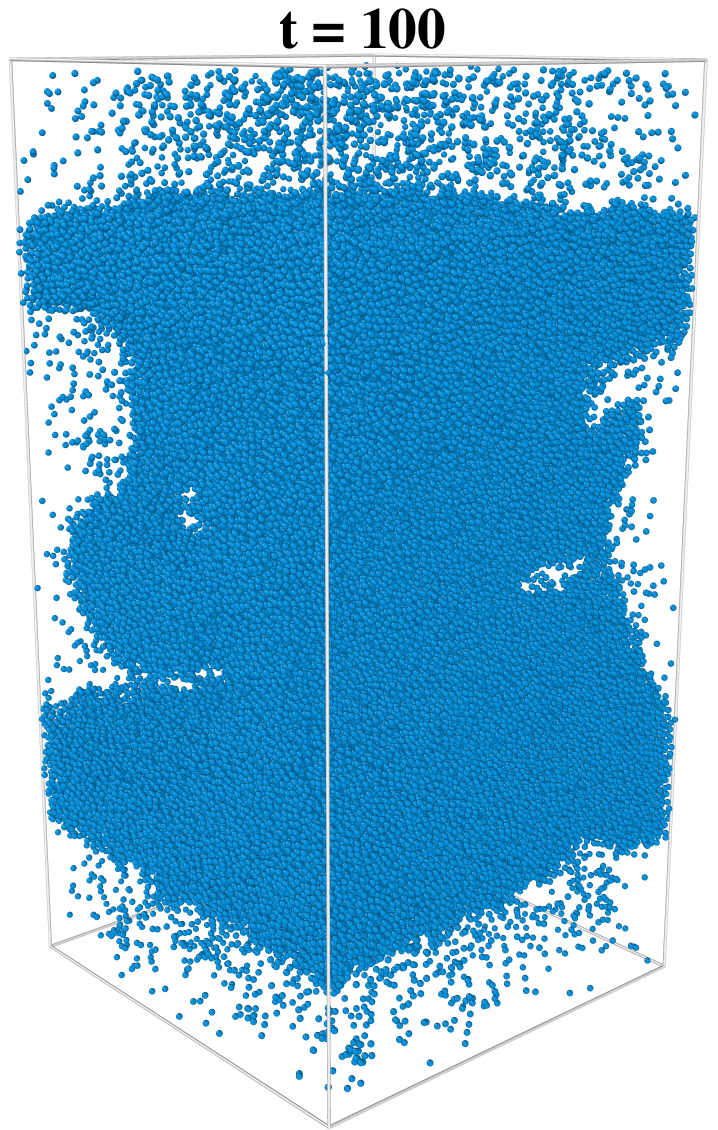}
    \end{subfigure}
    \begin{subfigure}
        \centering
        \includegraphics[width=0.45\linewidth]{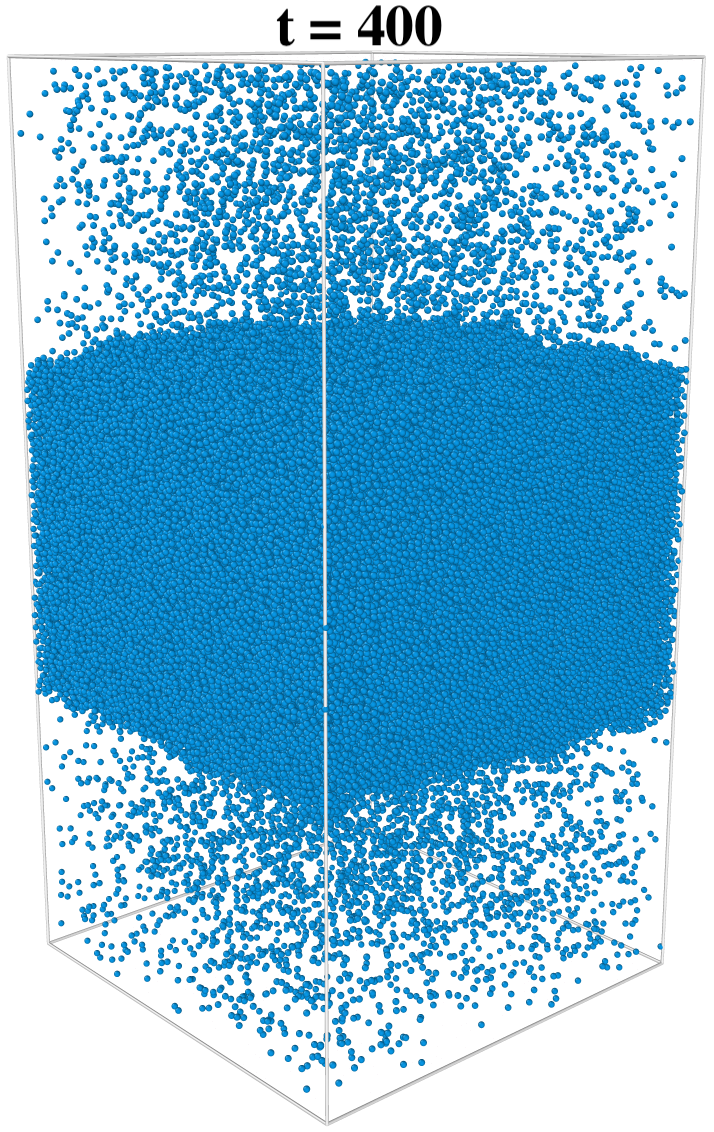}
    \end{subfigure}
    \caption{The representative snapshots of our phase separating vapor-liquid system at $T=0.8$ in the absence of external gravitational field. The blue dots represent the particles.}
    \label{fig1:zero_gravity}
\end{figure}
 
The system is first equilibrated at a high temperature of $T=10$  to get a homogeneous configuration of uniform density. It is then quenched to $T=0.8$, well below the critical temperature $T_c$ of the vapor-liquid transition to obtain the vapor-liquid system. For the study of vapor-solid phase separation, the system is quenched to $T=0.1$. A spatially uniform external gravitational force $g$ is applied downward in the z-direction after the quench. 

Representative configurations obtained from the MD simulation with time, shown in Fig.~\ref{fig1:zero_gravity} demonstrate the evolution morphology of the vapor-liquid system under consideration. The formation of percolating domains clearly suggest that the system is near the critical density. As discussed in the previous section, the standard parameter to analyze the phase ordering dynamics is the average lengthscale $\ell(t)$. To compute $\ell(t)$,  we introduce the two point equal-time correlation function defined as \begin{equation}
    \label{eq:correlation}
     C(\vec r,t)=\langle\psi(0,t)\psi(\vec r,t)\rangle - \langle\psi(0,t)\rangle\langle\psi(\vec r,t)\rangle 
 \end{equation}
 where the angular brackets denote statistical averaging and $\psi(\vec r,t)$ is the order parameter calculated by mapping the system to Ising model. $\psi(\vec r,t)$ is assigned the value +1 if the local density over a box size of $(2\sigma)^3$ located at $\vec r$ is greater than the critical density $\rho_c$ and -1 otherwise \cite{Rounak1,Rounak2}. 
The lengthscale $\ell(t)$ is estimated from the first zero value of the correlation function $C(r=\ell,t)=0$. Of course, there are other methods to calculate the domain size of the system like inverse of first moment of the structure factor, domain size distribution of the system, etc. The Fourier transform of the correlation function gives the structure factor given by
 \begin{equation}
\label{eq:structure_factor} 
   S(\vec k,t)= \int d\vec r\ exp(i\vec k.\vec r)\ C(\vec r,t) 
 \end{equation}
The presence of external gravitational force results in the anisotropic growth of the domains in different spatial directions. Hence the correlation function is calculated separately in each spatial direction \cite{daniyaPRE} rather than being spherically averaged.
\section{Results}
In this section we discuss the results for both the vapor-liquid and vapor-solid phase separating systems influenced by gravity.

\subsection{Vapor-Liquid phase separation}

We focus on the isothermal vapor-liquid phase separation triggered by a sudden thermal quench below the critical temperature $T_c$ in the presence of an external gravitational field. We begin our analysis with the zero gravity case which is considered as a reference point. In Fig.~\ref{fig1:zero_gravity} we show the time evolution of the system at $T=0.8$, near the critical density $\rho=0.3$ without any external field ($g=0$). During the coarsening process, the system exhibits a bi-continuous domain morphology. Eventually, a complete transition into vapor phase and liquid phase is observed. Note that the order parameter $\psi(\vec r,t)$ is the prime measure to distinguish between the vapor and liquid phase.
\begin{figure}[ht]
    \centering
    \begin{subfigure}
        \centering
        \includegraphics[width=0.425\linewidth]{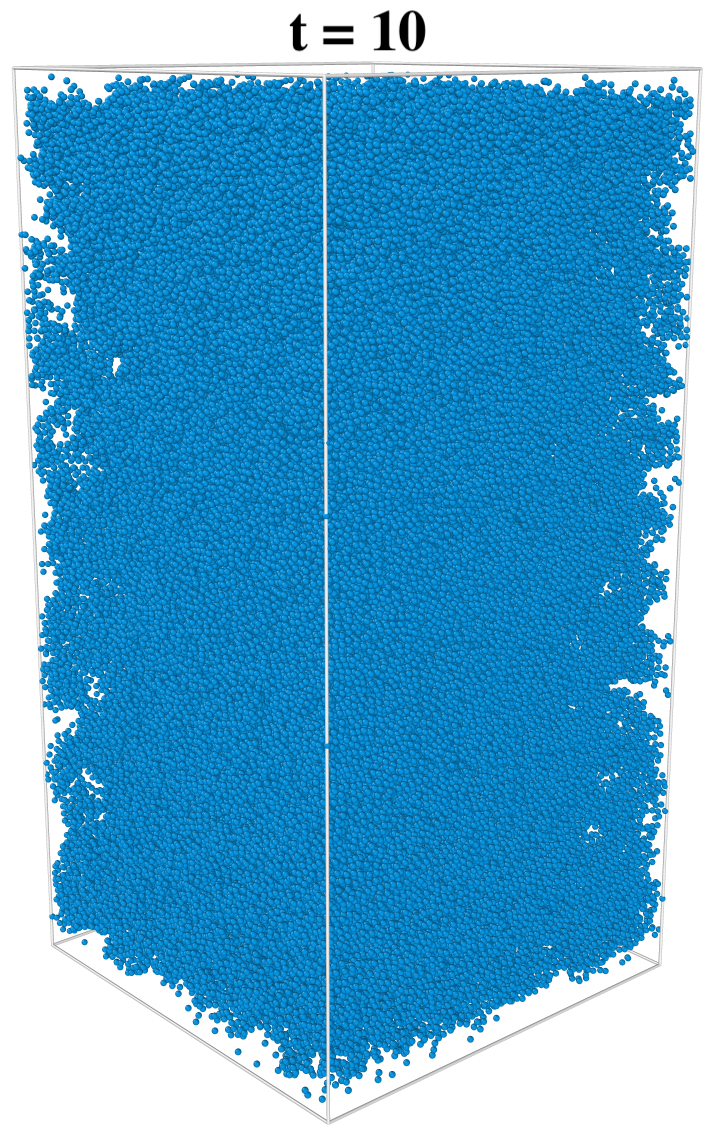}
     \end{subfigure}
    \begin{subfigure}
        \centering
        \includegraphics[width=0.425\linewidth]{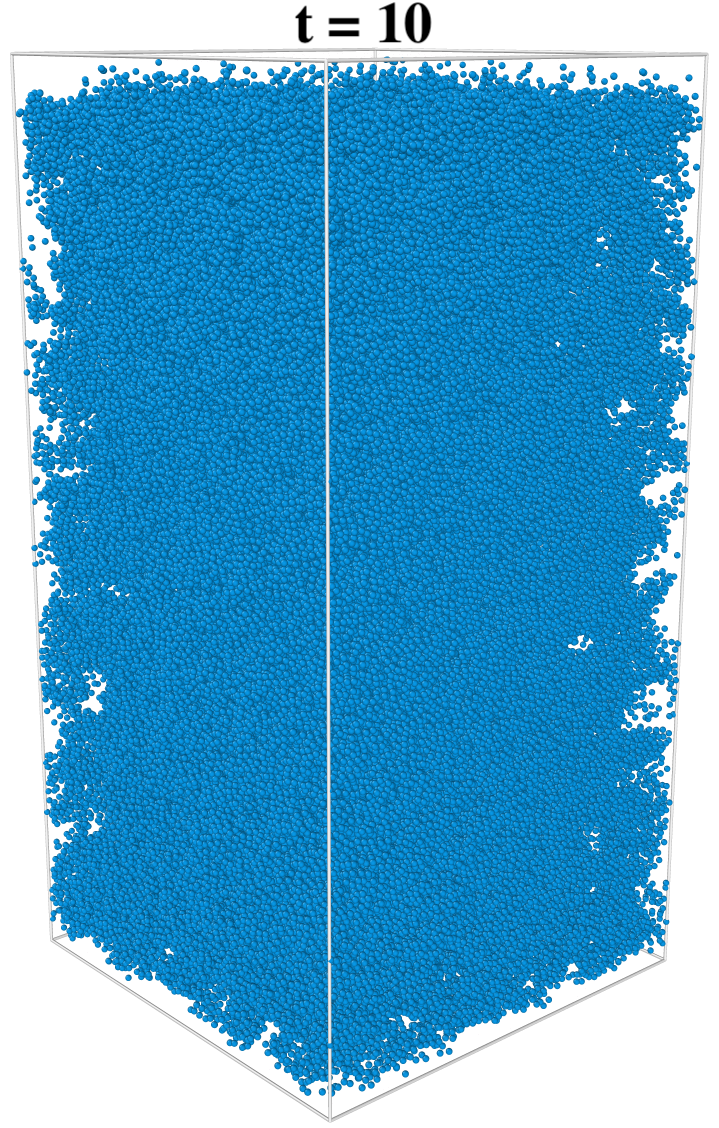}
    \end{subfigure}\\
    \begin{subfigure}
        \centering       
         \includegraphics[width=0.425\linewidth]{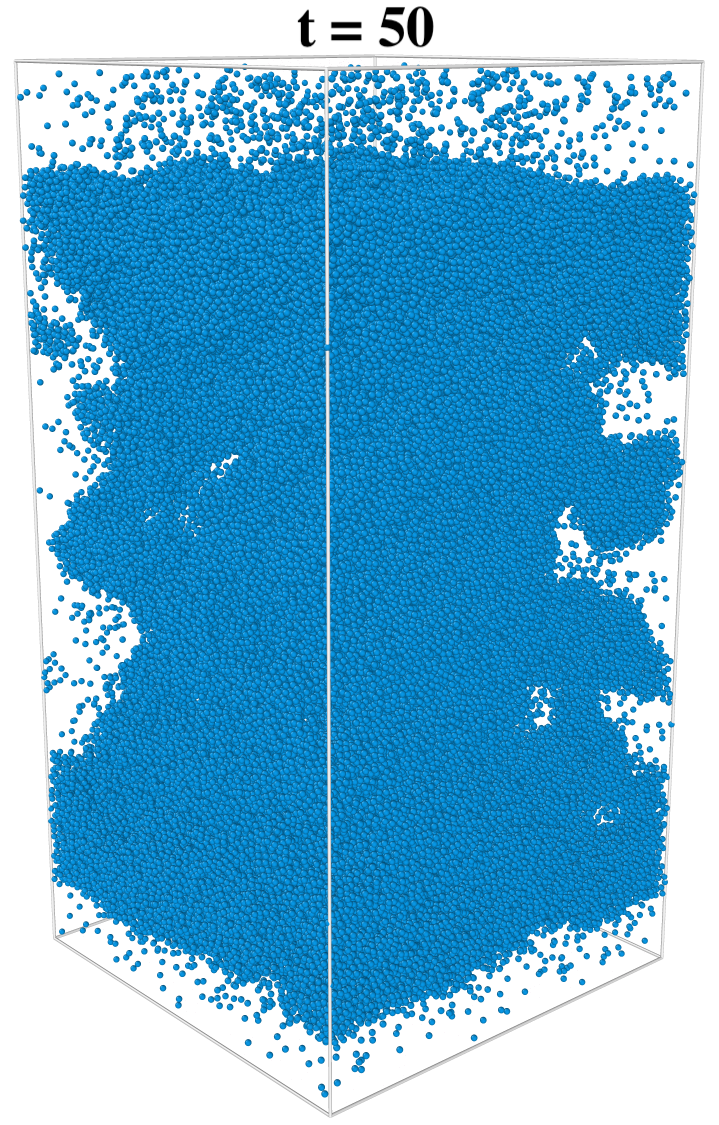}
    \end{subfigure}
    \begin{subfigure}
        \centering
        \includegraphics[width=0.425\linewidth]{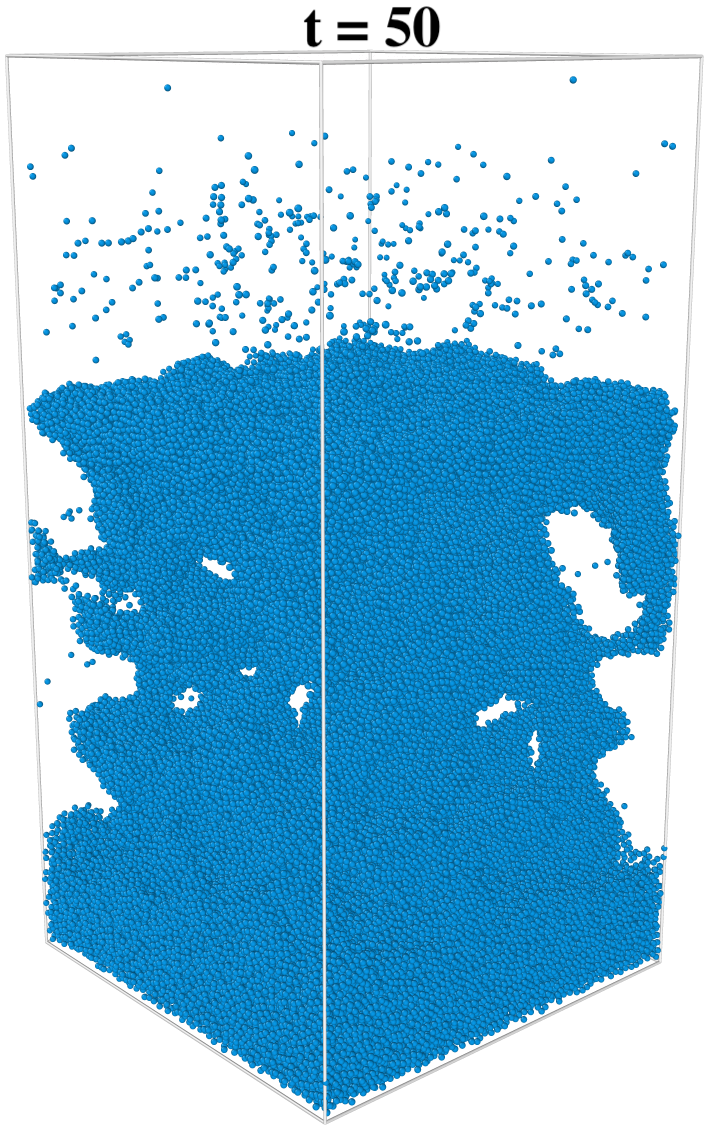}
    \end{subfigure}\\
    \begin{subfigure}
        \centering
        \includegraphics[width=0.425\linewidth]{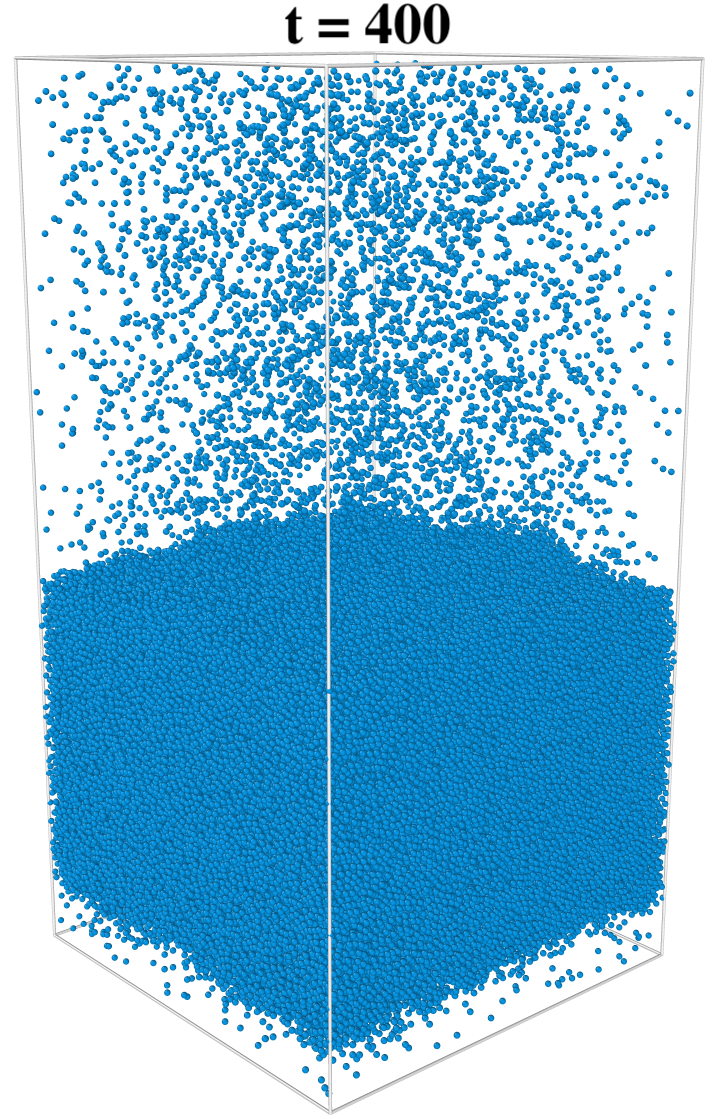}
     \end{subfigure}
    \begin{subfigure}
        \centering
        \includegraphics[width=0.425\linewidth]{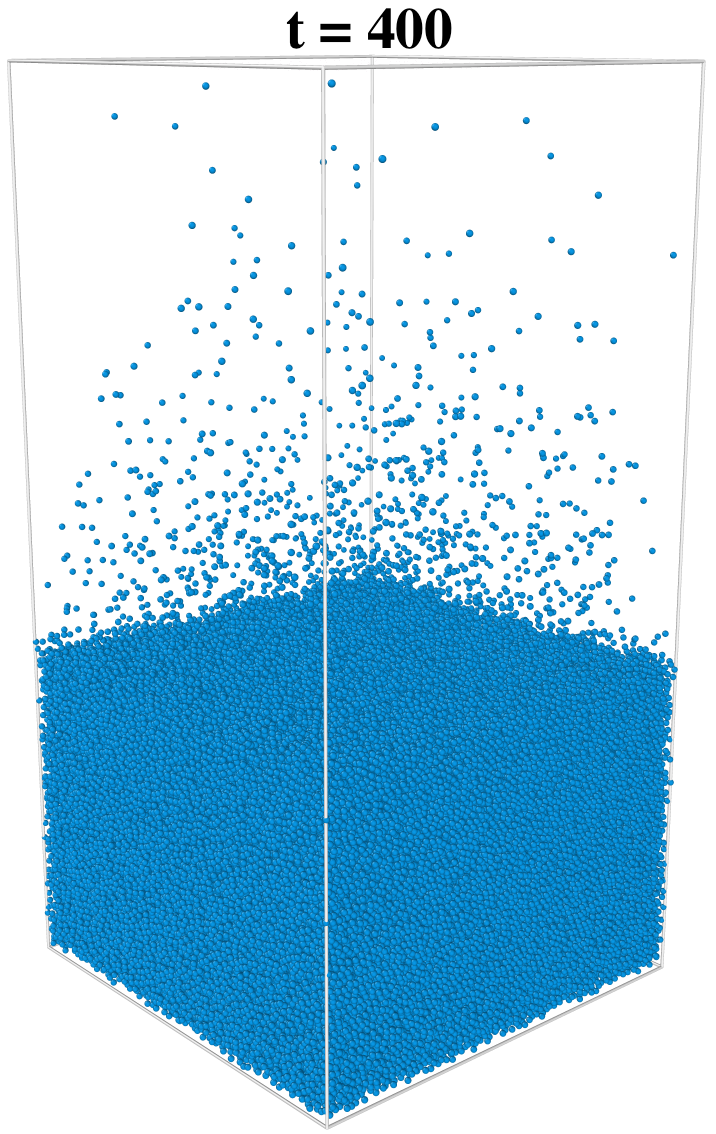}
    \end{subfigure}
    \caption{The morphological evolution of the vapor-liquid system with time under the influence of gravity value $g = 0.003$ is shown in the left column. In the right column, we show the same results for $g=0.05$.}
    \label{fig1:low&highgravity}
\end{figure}
   
Next we investigate the effect of gravity on the coarsening process of our system. Results are presented for a chosen set of $g$ values in the range (0.003, 0.05). Fig.~\ref{fig1:low&highgravity} shows the representative snapshots for the domain evolution of the systems for $g=0.003$ and 0.05. From Fig.~\ref{fig1:low&highgravity} it is conspicuous that the onset of phase separation occurs immediately after quenching, and the domains evolve with time. Here, along with surface tension and viscous forces, the gravitational field also plays an important role in determining the domain morphology and coarsening behavior. Due to coarsening process, particle rich liquid phase and particle deficit vapor phase grow with time. This facilitates the gravitational force to overcome the interfacial tension, and the denser phase moves downward due to gravity, analogous to sedimentation. Eventually a single interface between the lighter and heavier phase is formed.

A systematic change in the rate of structural evolution of the system is observed  with increasing gravitational force. At low $g$ values the rate of phase separation as well as sedimentation is much slower. This is finely evident in Fig.~\ref{fig1:low&highgravity} for $g=0.003$. Contrary to this, for higher $g$ values, the system phase separates at a faster rate and the denser phase accumulates at the bottom of the simulation box by creating completely distinguishable liquid and vapor domains.

\begin{figure}[ht]
	\centering
	\begin{subfigure}
		\centering
		{\includegraphics[width=0.9\linewidth]{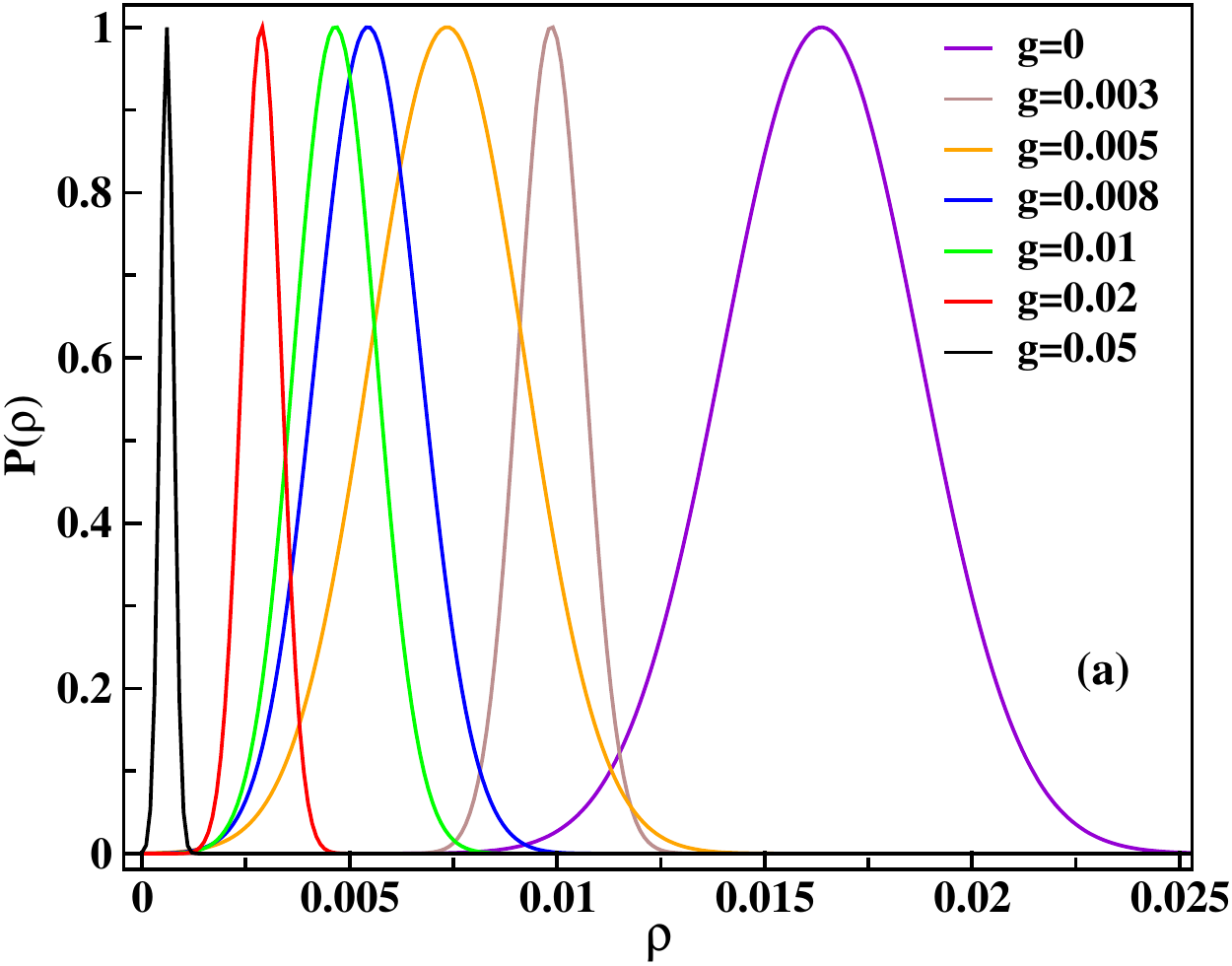}}
	\end{subfigure}\\
	\begin{subfigure}
		\centering
		{\includegraphics[width=0.9\linewidth]{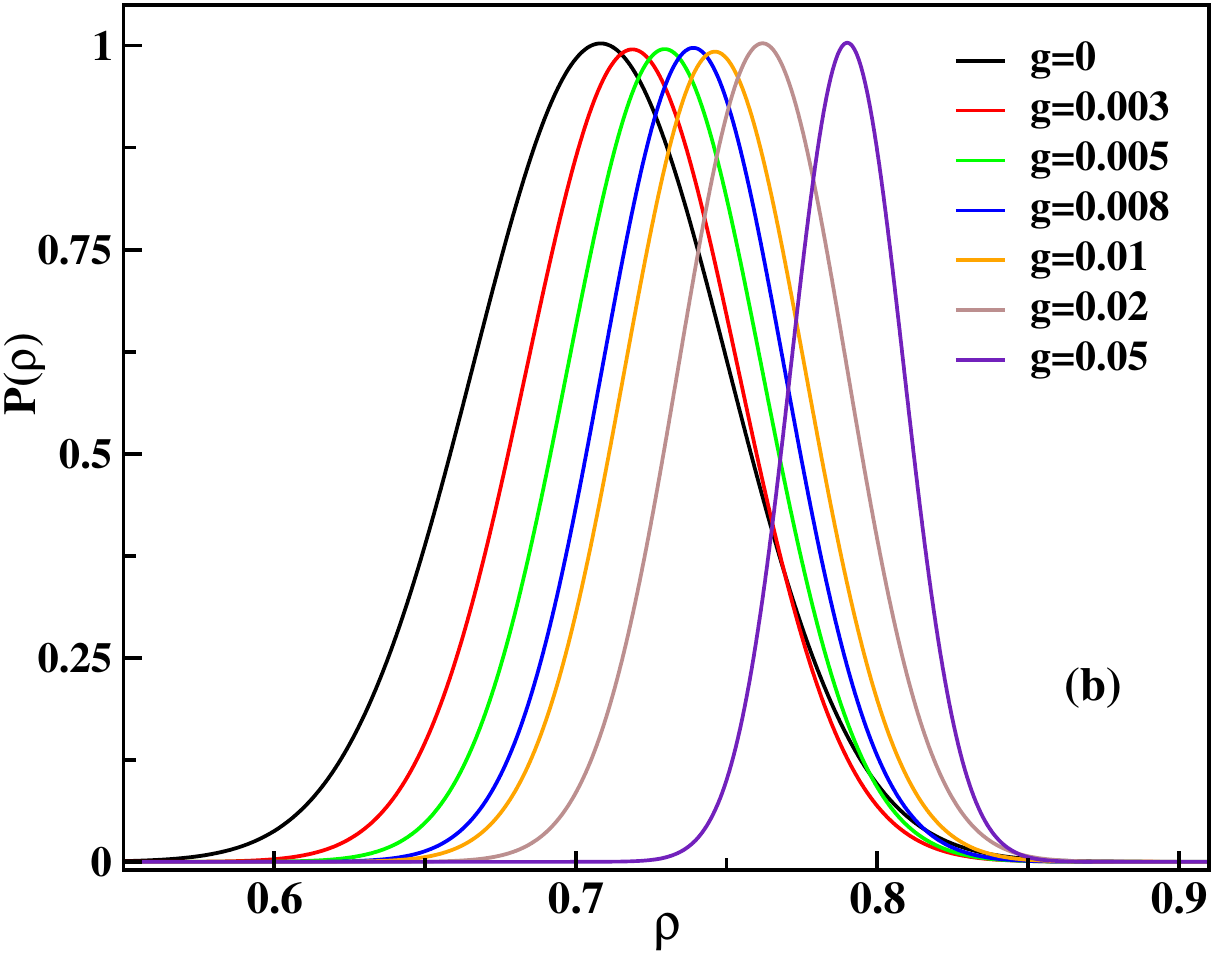}}
	\end{subfigure}
	\caption{The density distribution profile for the fully separated vapor and liquid phases are shown in (a) and (b) respectively.}
	\label{fig3:L_V_rho}
\end{figure}

It is also evident that the completely separated phases at long time are affected by the external field. For better understanding we calculate the density of the vapor and the liquid phase separately.  In Fig.~\ref{fig3:L_V_rho} we show the distribution of the densities P($\rho$) for the fully separated vapor and liquid phase for all $g$ values. For better visualization, the peak of the distributions is normalized to 1. The peak position of the distribution represents the average density. When the strength of the gravitational field is low, the vapor and liquid phases are closer to the zero gravity case. But as the field strength in increased, we observe the density of the vapor phase gradually decreases and the liquid phase becomes tightly packed. The average vapor and liquid phase density for the $g=0$ case is 0.02 and 0.71 respectively. When the gravitational field is present on the system, all the particles experience an additional force towards the bottom. As the strength of the field is increased, the gravitational pull becomes stronger and hence more particles are attracted towards the bottom and become the part of liquid phase making the domain denser. For $g=0.05$, which is the highest gravitational strength chosen in our simulation, the average vapor and liquid phase density are found to be 0.0001 and 0.8 respectively.

\begin{figure}[ht]
    \centering
    \begin{subfigure}
        \centering
        {\includegraphics[width=0.9\linewidth]{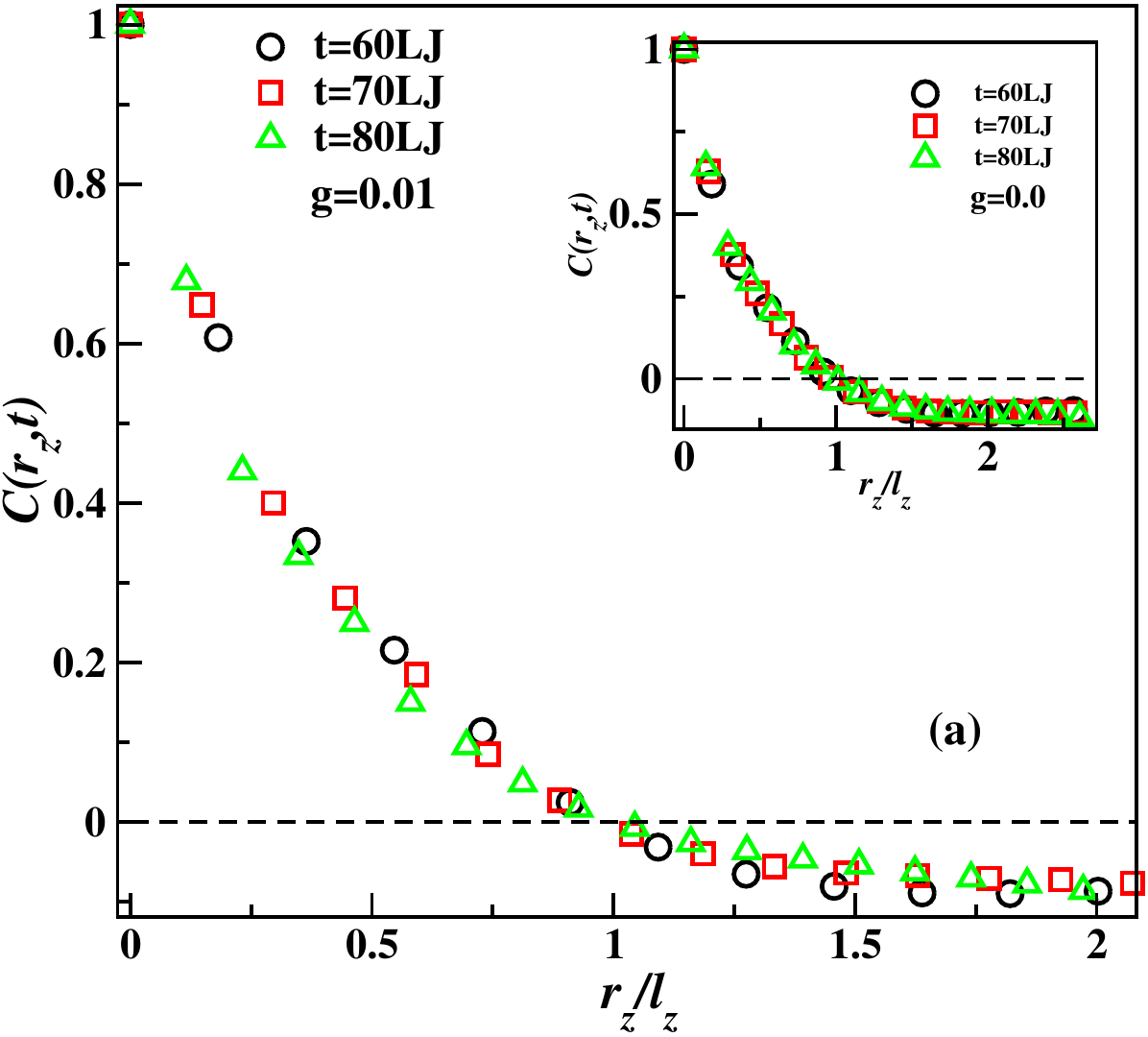}}
     \end{subfigure}\\
    \begin{subfigure}
        \centering
        {\includegraphics[width=0.9\linewidth]{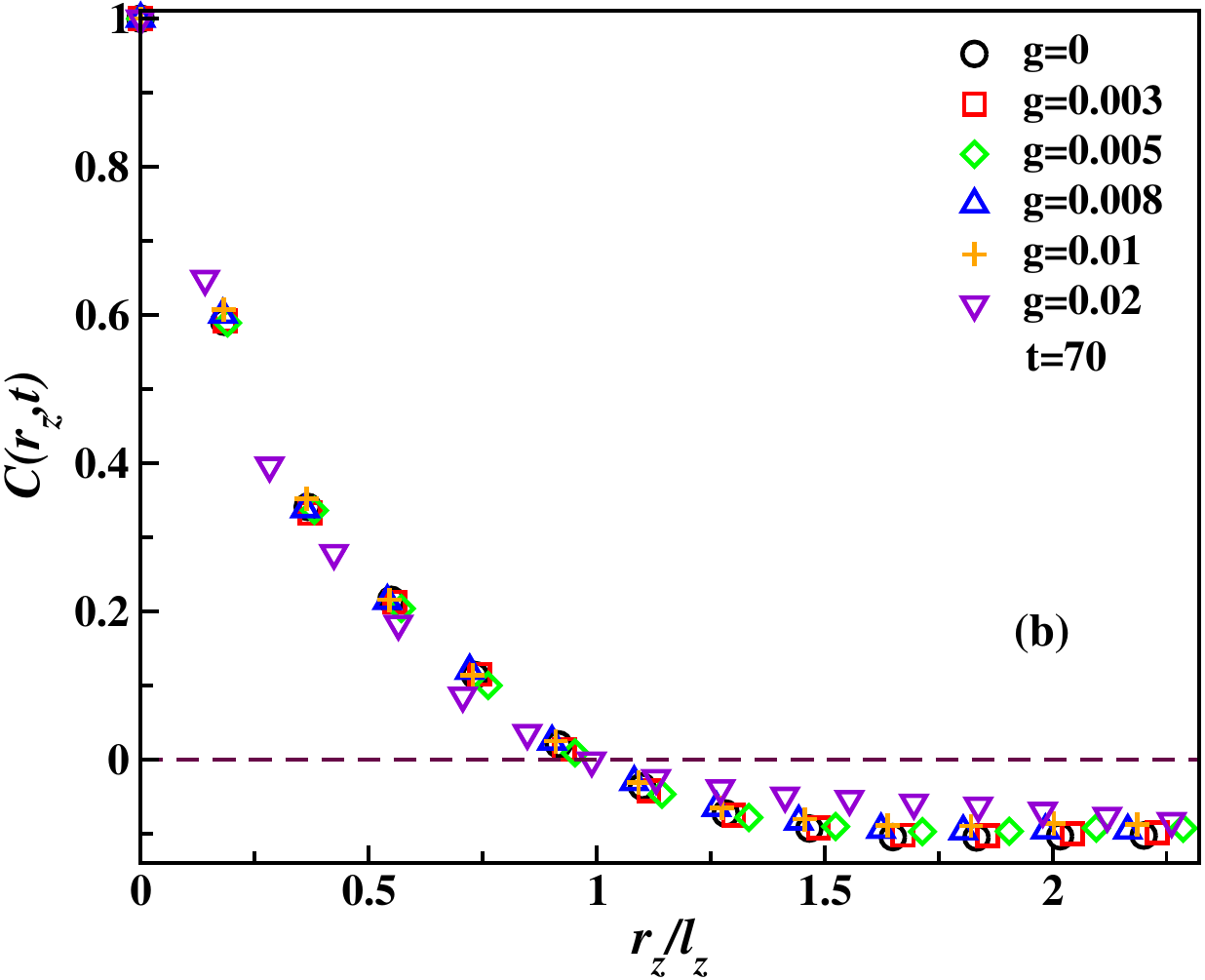}}
    \end{subfigure}
    \hfill
    \caption{(a) The scaling plots of $C(r_z,t)$ vs $r_z/\ell_z(t)$ for $g=0.01$ and 0 at different times are shown in the main figure and inset respectively. (b) The same scaling plots are shown for different $g$ values at a fixed time.}
    \label{fig4:scaled_corr}
\end{figure}

To characterize the domain morphology and study the domain growth we compute the two-point equal time correlation function $C(r_z,t)$ along the direction of gravity, from Eq.~\ref{eq:correlation}. The same is plotted alongside with scaled length $r_z/\ell_z(t)$ in Fig.~\ref{fig4:scaled_corr}(a) for $g=0$ and 0.01. An excellent data collapse is observed for different times. This exercise is repeated for the other $g$ values (not shown) and the quality of overlap is found to be the same. This is the well known Porod law and remains vindicated in the presence of gravity \cite{Shaista}. This also suggests that even in the presence of gravity, the system belongs to the same dynamical universality class \cite{Binder2}. Further, we find the correlation functions corresponding to different $g$ values overlap with each other when plotted at a fixed time. This is shown in Fig.~\ref{fig4:scaled_corr}(b). Therefore, the superuniversality (SU) property applies to our system \cite{SU1, SU2}.

\begin{figure}[h]
 \centering
 {\includegraphics[width=0.48\textwidth]{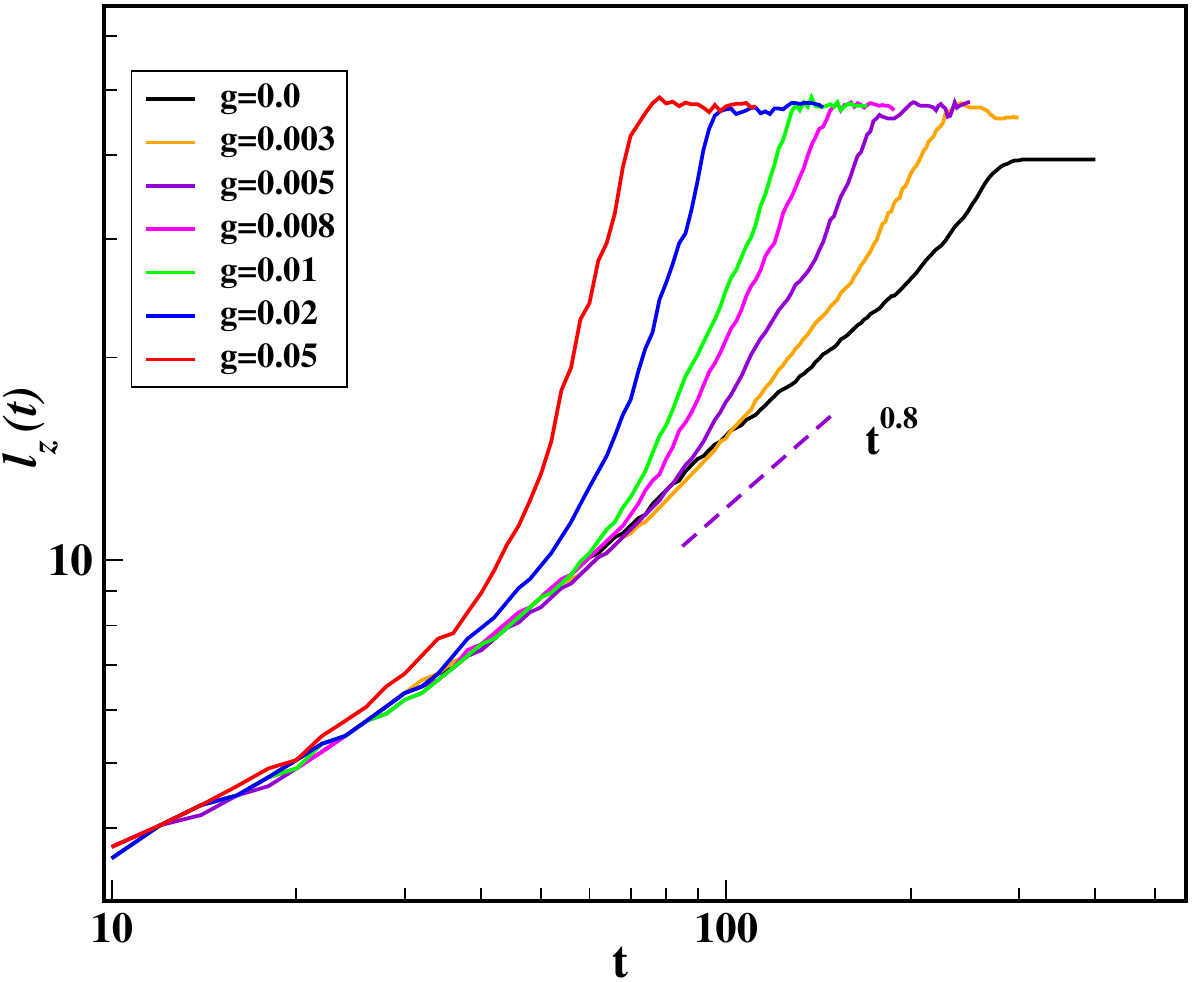}}
 \caption{ The time evolution of the average domain size $\ell_z(t)$ are shown for different $g$ values. The dashed line represents the guideline for the slope.}
 \label{fig5:z_componentliquid}
\end{figure}

To gain a qualitative understanding of the growth of average domain size with time it is sufficient to calculate the length scale $\ell(t)$. Due to the expected anisotropic growth in the presence of gravity, we compute the length scale separately in all three directions. We begin with the z-component of the length scale, $\ell_z(t)$ along which the gravitational field is applied. The time dependence of $\ell_z(t)$ is shown in Fig.~\ref{fig5:z_componentliquid} for different $g$ values. Our simulation is able to access the viscous hydrodynamic regime after an initial transient period. In the absence of gravity $(g = 0)$ we expect a linear behavior of $\ell_z(t )$. The slight deviation from this can be attributed to the nonzero offsets at the crossovers, which can be subtracted from $\ell(t )$ to recover the proper linear growth \cite{Das}. A dramatic change in the growth behavior is observed under the influence of gravity. At the early stage and $g>0$, the $\ell_z(t )$ exhibits linear growth, typical of the viscous regime and independent of $g$ value, suggesting that a universal scaling exists in the gravity direction. After that, a crossover to a considerably faster growth ($\alpha>1$) regime is observed. This crossover happens at an earlier time for the systems with higher $g$ value. This is consistent with the observations in Fig.~\ref{fig1:low&highgravity}. The fastest growth rate observed in our simulations corresponds to $\alpha=3$. 

This accelerated domain growth in the presence of external drive can be comprehended as follows. After the initial coarsening the liquid domains become moderately larger and heavier. A transition to gravity induced flow arises when heavy domains become unstable, i.e. the gravitational force on the denser phases overcomes the interfacial tension that keeps them suspended. This phenomenon can be likened to sedimentation. Once the instability sets in, additional length scales emerge due to accelerated growth in the direction of gravity.

\begin{figure}[ht]
	\centering
	\begin{subfigure}
		\centering
		{\includegraphics[width=0.8\linewidth]{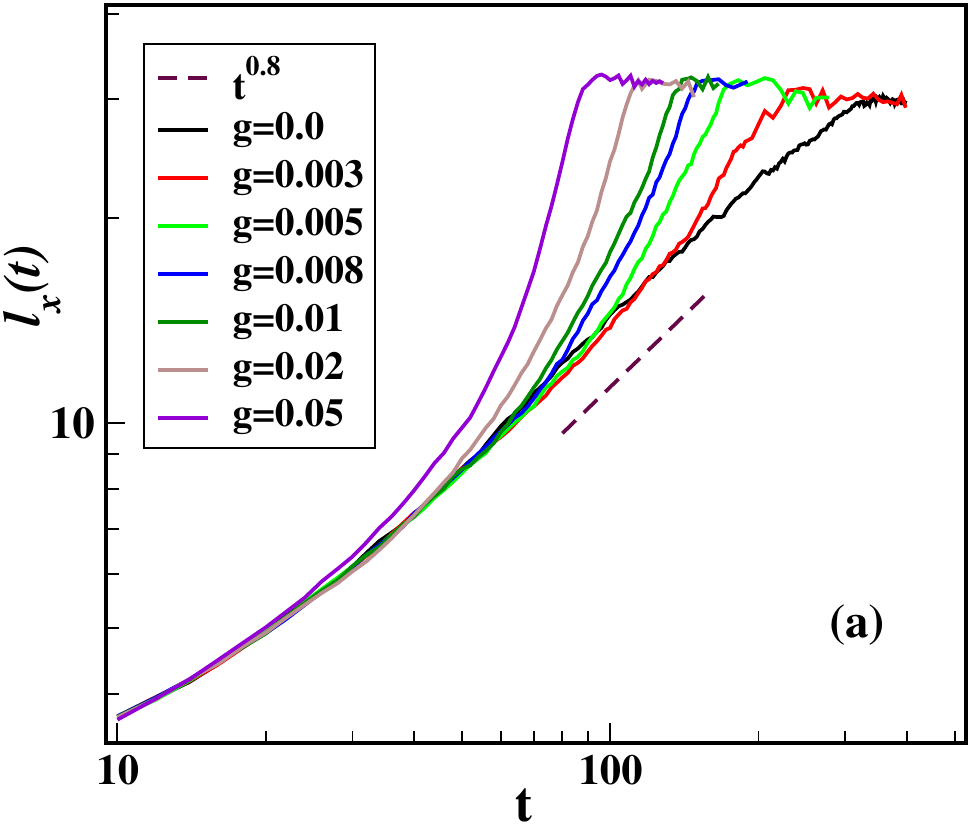}}
	\end{subfigure}\\
	\begin{subfigure}
		\centering
		{\includegraphics[width=0.8\linewidth]{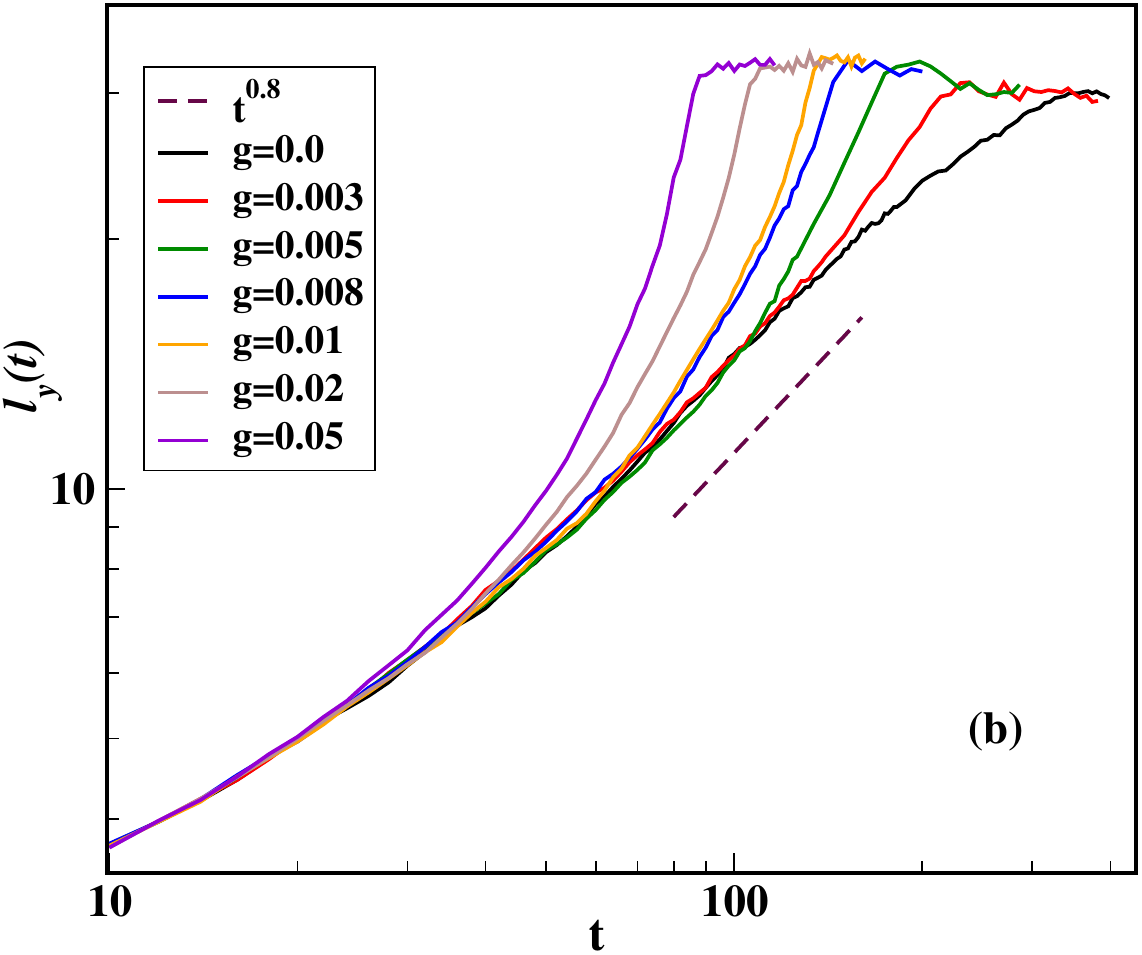}}
	\end{subfigure}
	\caption{The time evolution of the average domain size $\ell_x(t)$ and $\ell_y(t)$ for different $g$ values are shown in (a) and (b) respectively. The dashed lines represent the guideline for the slope.}
	\label{fig6:l_scaled}
\end{figure}

Next we focus on the characteristic domain growth in the horizontal directions. Due to symmetry in the system, the lengthscale in the x and y directions is expected to analogous. These results are shown in Fig.~\ref{fig6:l_scaled}. For the $g=0$ case, the growth follows approximately a linear behavior. In the presence of gravity ($g>0$), again we see an accelerated growth ($\alpha>1$) after a transient period.  

\begin{figure}[ht!]
	\centering
	{\includegraphics[width=0.4\textwidth]{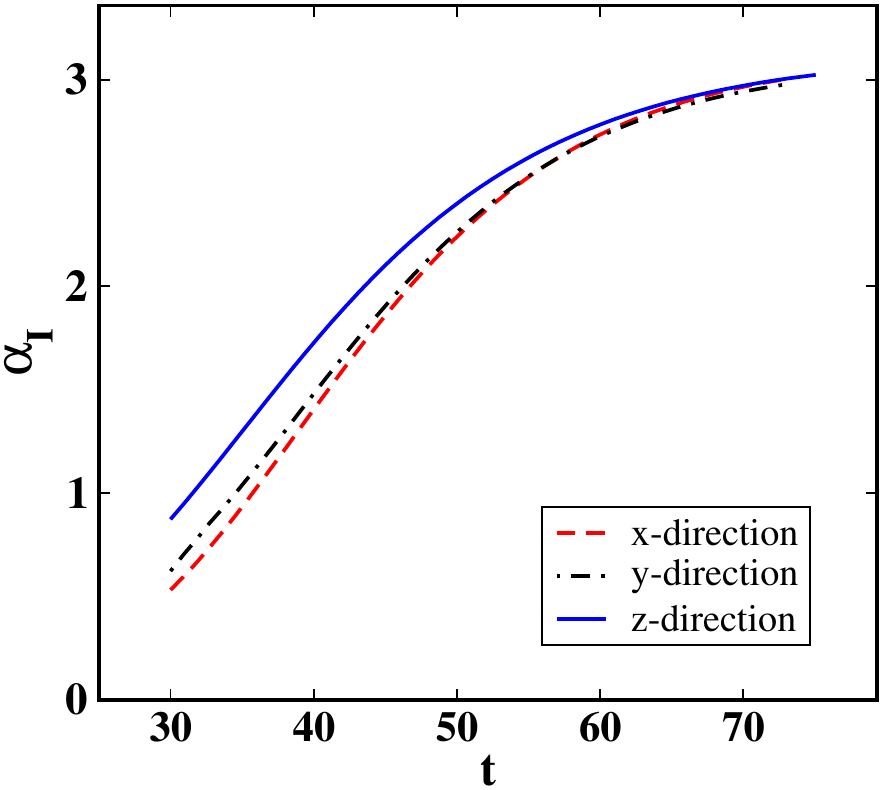}}
	\caption{The instantaneous slope of the length scale vs time for $g=0.05$.}
	\label{fig7:slope}
\end{figure}

To gain a comprehensive understanding of the relative growth rate in different directions, we compute the instantaneous slope of the length scale defined as $\alpha_\mathrm{I} = d(\mathrm{ln}\ell(t)/d(\mathrm{ln}t)$ for the highest gravity value of $g=0.05$ \cite{Saikat}. The results are shown in Fig.~\ref{fig7:slope}. The exponent is found to be higher in the z direction at the intermediate time. Therefore, growth rate is faster along the direction of gravity and the system is anisotropic. Over the interim period, the exponent experiences a gradual increase. This phenomenon can be attributed to the expanding size of the liquid domains over time, wherein the gravitational force becomes increasingly influential, thereby prompting accelerated growth. At late time, the growth exponent converges to the value $\alpha \sim 3.0$ for all the directions and the system restores the isotropic nature. 

The intermediate accelerated growth rate in the z direction can be understood in the following way. Subsequent to the initial phase separation period, denser domains commence precipitating at the lower portion of the simulation box. This circumstance leads to a faster growth rate in the vertical direction compared to the horizontal direction. As the sedimentation process nears its conclusion, the liquid domains spread horizontally, thereby hastening the growth of domains along the x and y directions. Eventually, the growth rate becomes uniform in all dimensions and the system regains its isotropic characteristics.

\begin{figure}[ht!]
	\centering
	{\includegraphics[width=0.4\textwidth]{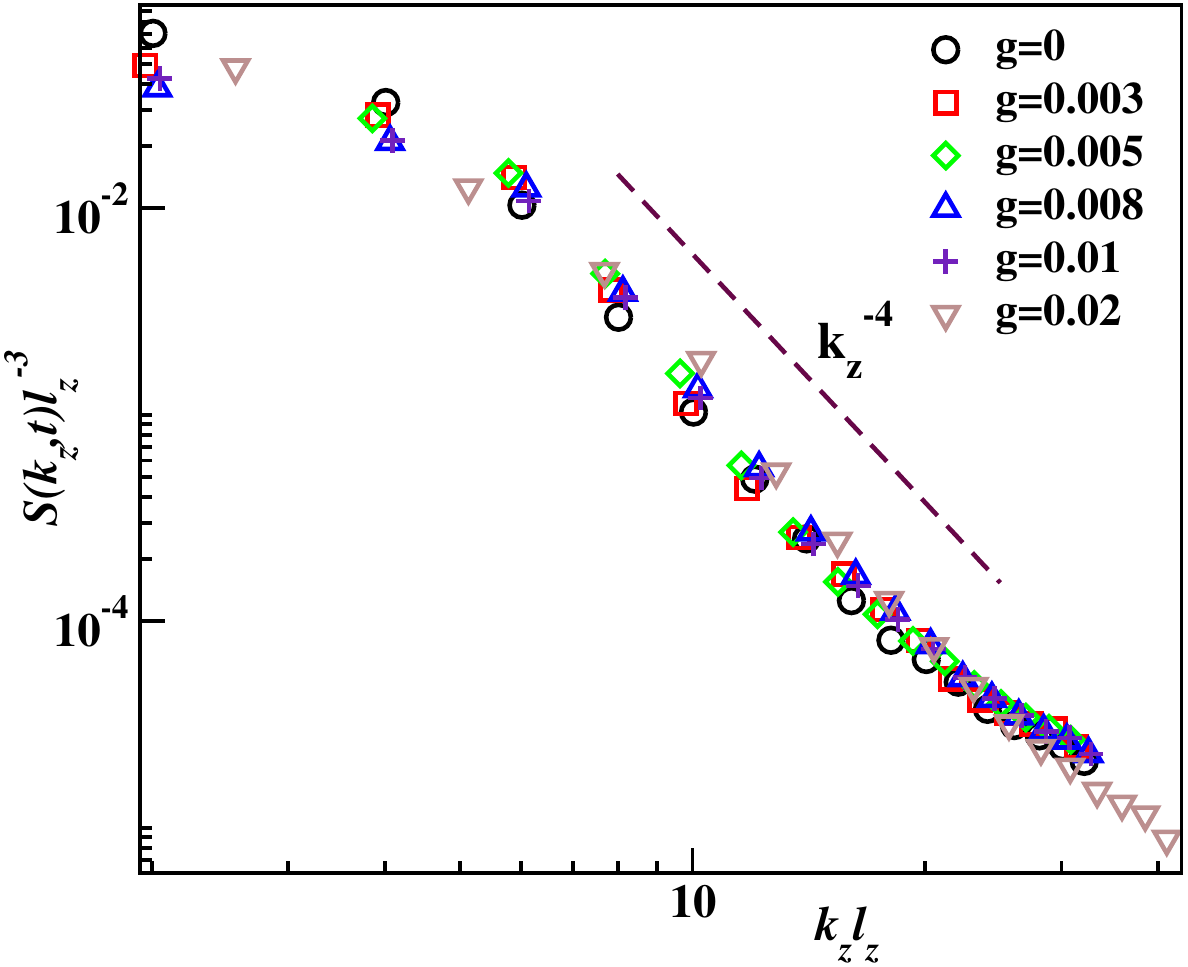}}
	\caption{The scaled structure factor $S(k_z,t)\ell_{z}^{-3}$ vs $k_z\ell_z$ is plotted for different $g$ values. The straightline is a guideline for the Porod law $S(k_z) \sim k^{-4}$.}
	
	\label{fig8:structure_factor}
\end{figure}

For further quantification of the structural details we focus on the structure factor given by Eq.~\ref{eq:structure_factor}. The structure factor $S(k,t)$ is has a scaling form,
\begin{equation}
    \label{eq:scaled_structure}
     S(\vec k,t) \equiv \ell^d \tilde S (\vec k\ell(t))
 \end{equation}
 in systems with self-similar patterns where $\tilde S$ is another time-independent master scaling function. The power-law behavior observed in the tail of $S(\vec k,t)$  has the following form,
\begin{equation}
    \label{eq:porod_law}
      S(\vec k,t)\sim  k^{-(d+1)}
 \end{equation}
 where d is the dimension of the ordering system. The $k$-dependence of Eq.~\ref{eq:porod_law} is referred to as the Porod's law \cite{Binder2}. In Fig.~\ref{fig8:structure_factor} we show the scaled structure factor $S(k_z,t)\ell_z^{-3}$ vs $k_z\ell_z$ for different $g$ values. We can see an agreeable data collapse from the figure. The dashed line represents the power law $k_z^{-4}$. The strong agreement of our large $k_z$ data with the Porod's law is clearly evident. These results are consistent with the observations in Fig.~\ref{fig4:scaled_corr}b and substantiate the validation of superuniversality. 

\subsection{Vapor-Solid phase separation}
\begin{figure}[ht]
    \centering
    \begin{subfigure}
        \centering
        \includegraphics[width=0.45\linewidth]{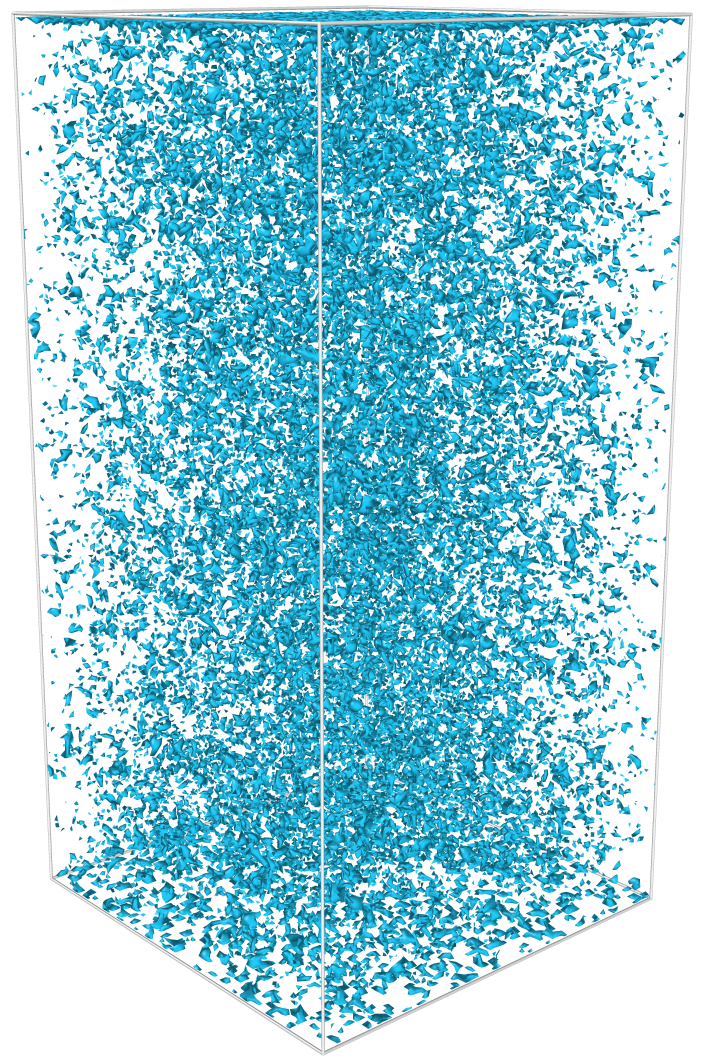}
     \end{subfigure}
    \begin{subfigure}
        \centering
        \includegraphics[width=0.45\linewidth]{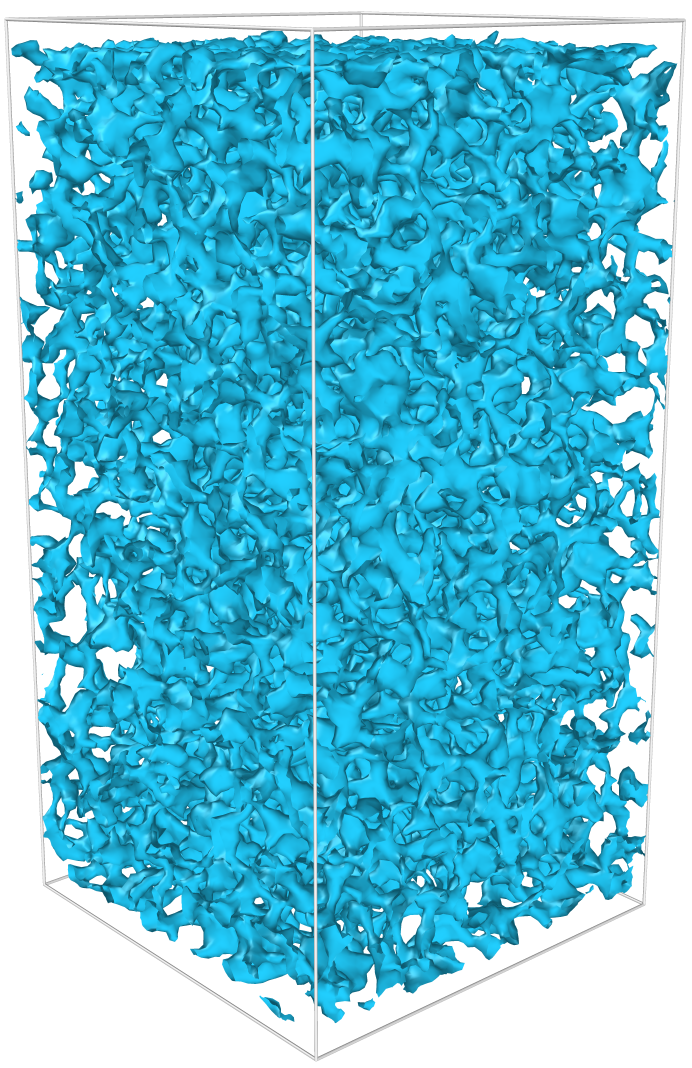}
    \end{subfigure}\\
    \begin{subfigure}
        \centering        
        \includegraphics[width=0.45\linewidth]{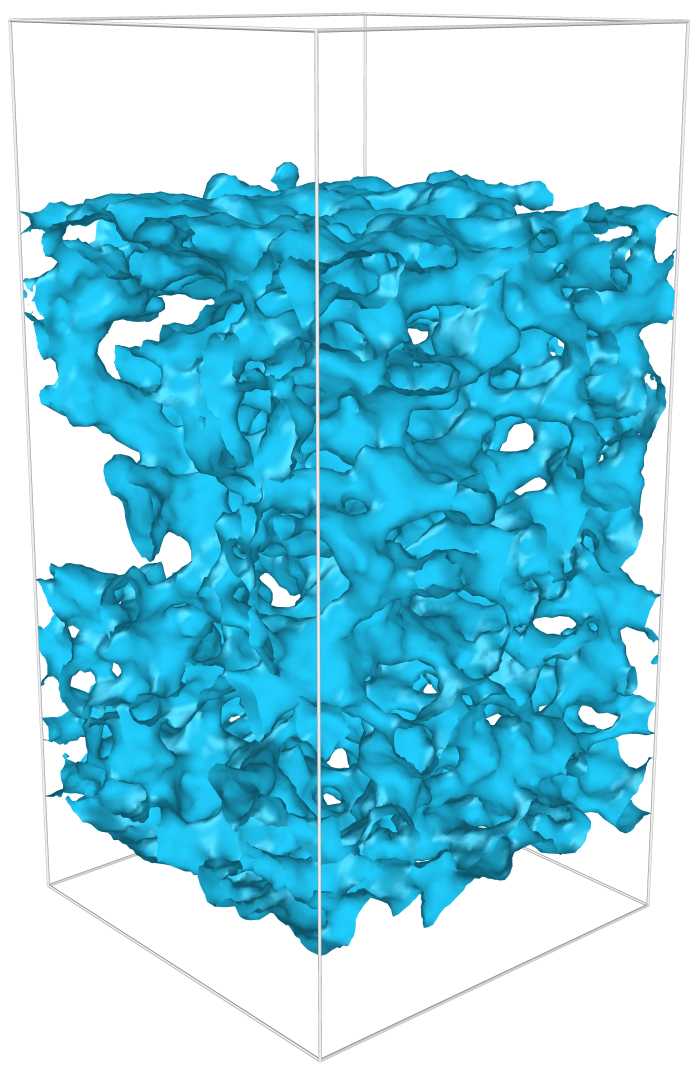}
    \end{subfigure}
    \begin{subfigure}
        \centering
        \includegraphics[width=0.45\linewidth]{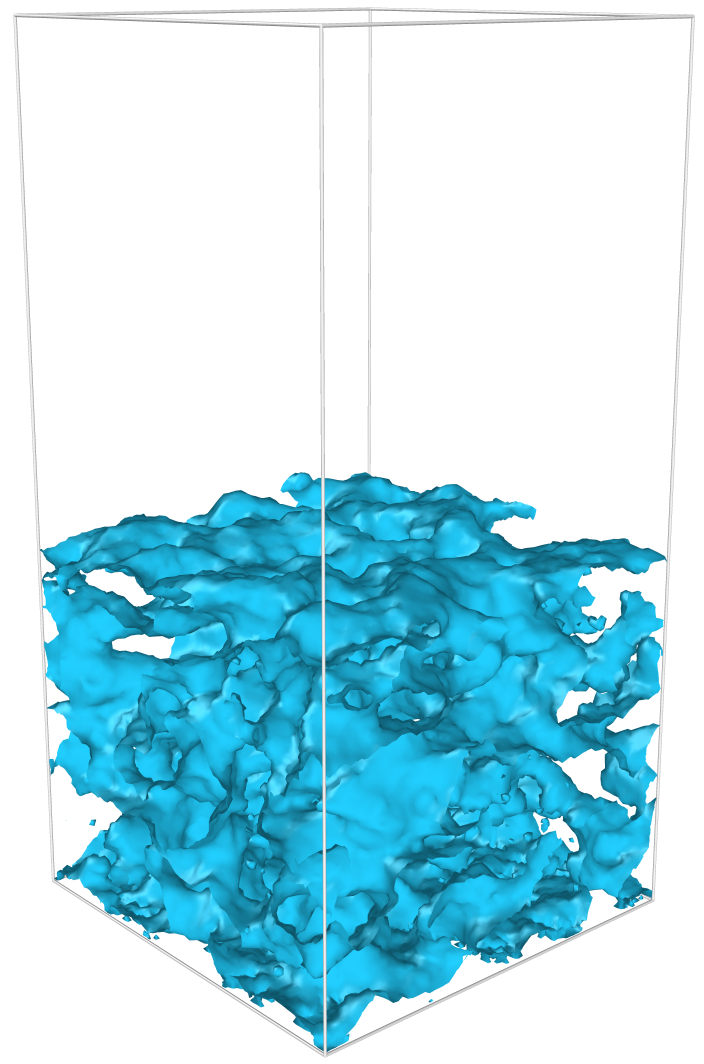}
    \end{subfigure}
    \caption{The representative snapshots for the vapor-solid phase separating system in the presence of external gravitational field of $g=0.003$.}
    \label{fig9:solid_gravity003}
\end{figure}

In this subsection we focus on the kinetics of vapor-solid phase separation under the influence of gravity. We begin the simulation by preparing a homogeneous system at high temperature of $T = 10$. The vapor-solid phase separation is initiated by quenching the system at density $\rho=0.3$ to a very low temperature $T = 0.1$ at time $t=0$. The sufficiently low temperature guarantees solid phase formation \cite{Saikat1}.  The system is then allowed to evolve via MD simulation. The time evolution of our phase separating system under the influence of a chosen gravitational field is shown in Fig.~\ref{fig9:solid_gravity003}. For better visualization of the solid phase formation, we show the surface mesh plot rather than the particles. The interconnected solid domains formation is clearly observed in the figure. The effect of gravity forces the domains to move downward. As we analyze the evolution snapshots under various gravitational strengths (not presented here) it is observed that the rate of sedimentation varies according to the applied field strength. 

A qualitative understanding of domain growth is obtained from the characteristic lengthscale $\ell(t)$. Due to the anticipated anisotropy in the system, the lengthscale is calculated separately in the gravity direction and the horizontal direction, as elaborated in the preceding section. The $\ell(t)$ is calculated from the domain length distribution in a particular direction, a conventional approach employed in the dynamics of phase ordering \cite{Roy1,Majumdar1,Roy3,Roy4}.

\begin{figure}[ht]
	\centering
	\begin{subfigure}
		\centering
		{\includegraphics[width=0.8\linewidth]{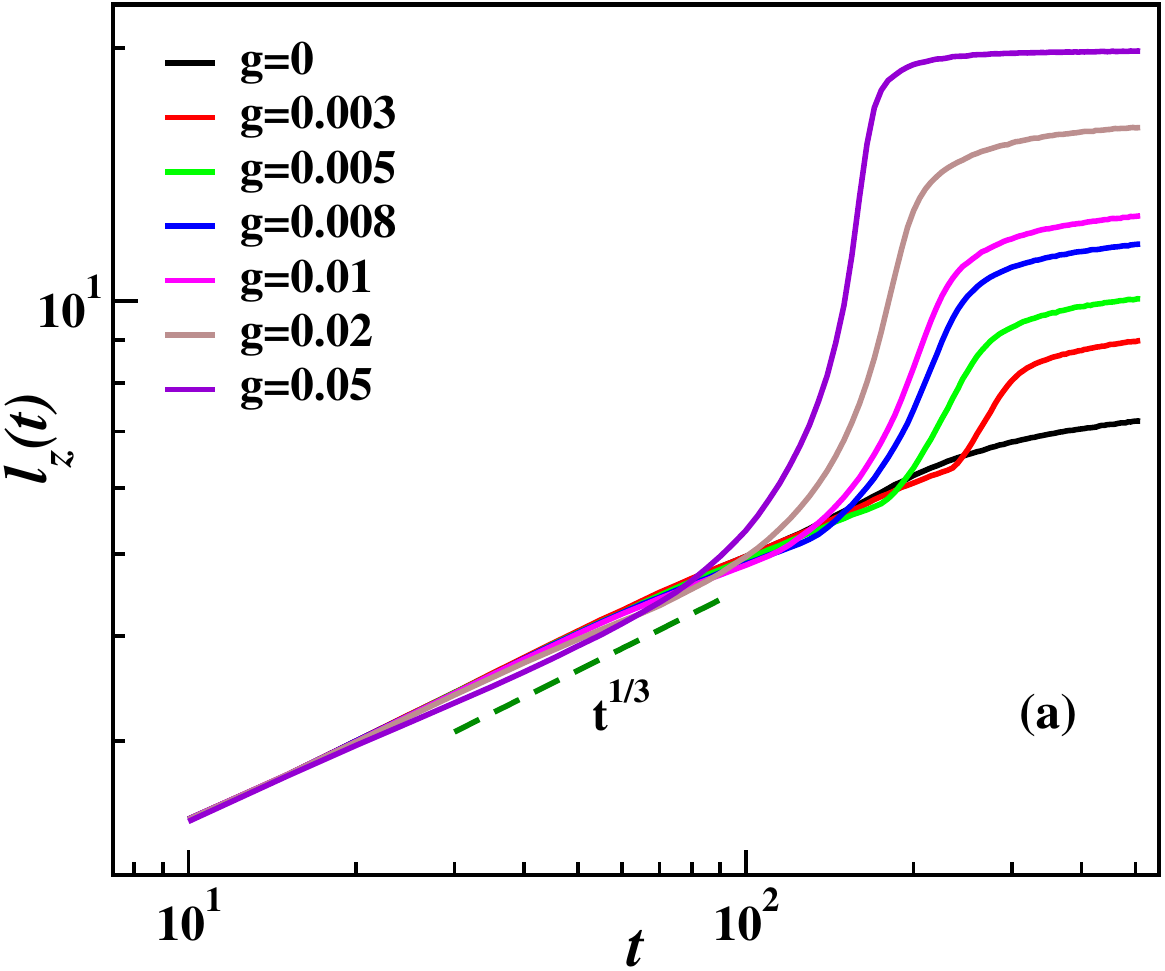}}
	\end{subfigure}\\
	\begin{subfigure}
		\centering
		{\includegraphics[width=0.8\linewidth]{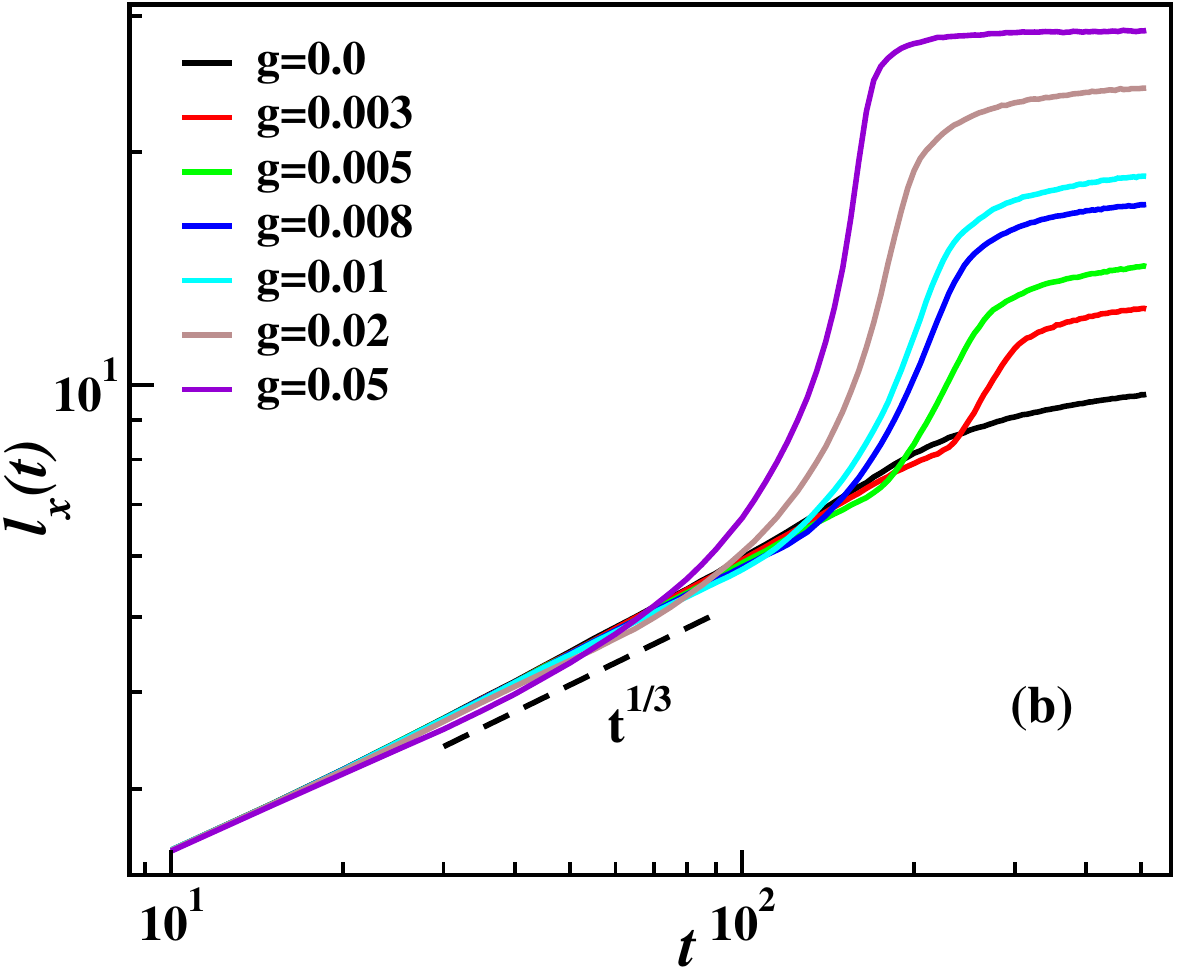}}
	\end{subfigure}
	\caption{The time evolution of the average domain size $\ell_z(t)$ and $\ell_x(t)$ for different $g$ values are shown in (a) and (b) respectively. The dashed lines represent the guideline for the slope.}
	\label{fig10:Solid_length}
\end{figure}

In Fig.~\ref{fig10:Solid_length}a we show the time variation of $\ell_z(t)$ in the z direction for a chosen set of $g$ values mentioned in the inset. According to Lifshitz-Slyozov law, in solid mixtures the spinodal decomposition takes place through the transport mechanism of diffusion and hence the growth exponent is $\alpha = 1/3$. In the absence of gravity, we observe that the growth exponent aligns with the Lifshitz-Slyozov law. However, when the influence of gravity comes into play, the growth dynamics undergo a dramatic change. Similar to vapor-liquid phase separation, following the initial transient period where the growth law is unaffected by gravity, a crossover to an accelerated growth is observed. The Fig.~\ref{fig10:Solid_length}a illustrates that the domain growth exponent exhibits a systematic variation with increasing field strength. The increase in the growth exponent stems from the collective motion of domains in the field direction, leading to the rapid growth of domains. 

An important point to note here, distinct from the previous section is that the saturation length scale  increases systematically. This can be comprehended as follows. The snapshots displayed in Fig.~\ref{fig9:solid_gravity003} clearly demonstrate that, over an extended period, the solid domains remain bicontinuous, and a complete phase separation into a single solid domain is never attained. On the other hand, there exists bicontinuous vapor phases as well, in between the solid domains. With increasing field strength the solid domains undergo a compressive force towards the bottom of the system. As a result the bicontinuous solid domains compactify and coalesce into a more interconnected, larger-sized domain. Consequently, with gravity the saturation length scale increases.

For the horizontal direction, we show the length scale in Fig~\ref{fig10:Solid_length}b in the x direction. Due to the obvious symmetry in the system, we abstain from showing results in the y direction. In the zero-gravity scenario, as anticipated, the growth exponent is found to be $\alpha=1/3$. Similarly, in the initial regime of finite gravity cases, the exponent is expected to remain the same. With time, as the domains increase in size, the influence of gravity becomes more pronounced. In the horizontal direction, the growth of the domain also escalates gradually, owing to the larger size of solid domains in the overall direction, not limited to the gravitational axis alone. Consequently, an accelerated domain growth is observed. 

\begin{figure}[ht!]
	\centering
	{\includegraphics[width=0.45\textwidth]{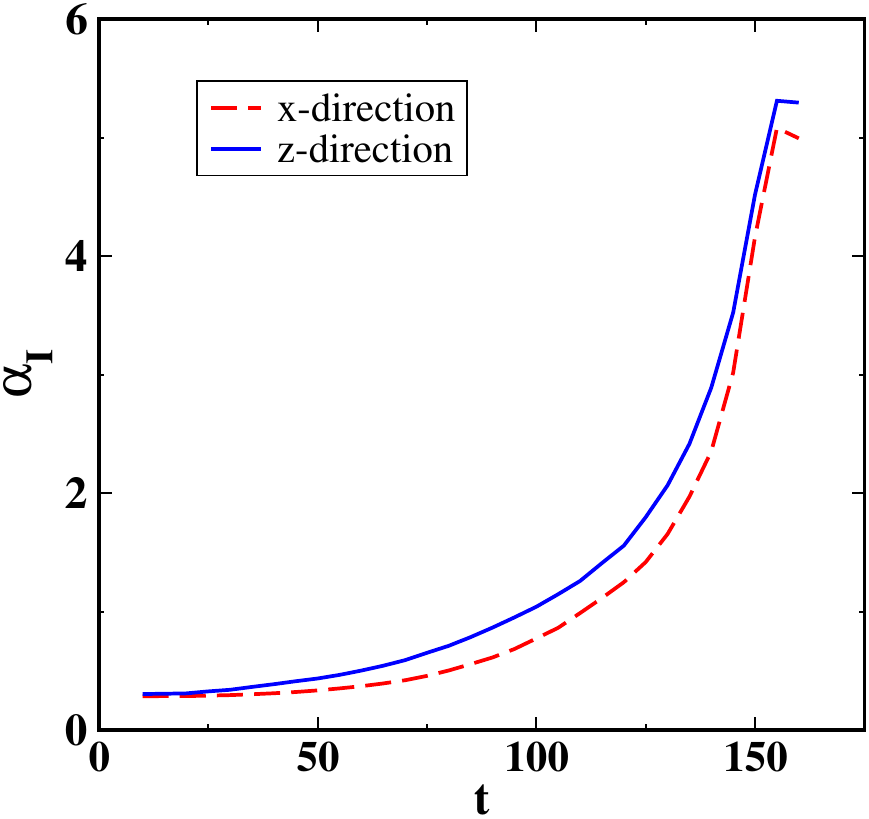}}\\
	\caption{The instantaneous slope of the length scale vs time for $g=0.05$.}
	\label{solid_insSlope}
\end{figure}

To quantify the growth rate, we show in Fig.~\ref{solid_insSlope} the instantaneous slope of the length scale $\alpha_\mathrm{I} = d(\mathrm{ln}\ell(t)/d(\mathrm{ln}t)$ associated with the highest field strength $g=0.05$.  A faster growth rate is discernible in the direction of gravity at the intermediate stage. It is noteworthy that the growth rate in this scenario surpasses that observed in vapor-liquid phase separation influenced by gravity. This disparity arises due to the increased weight of the solid domains compared to the liquid domains.

\begin{figure}[ht!]
	\centering
	{\includegraphics[width=0.45\textwidth]{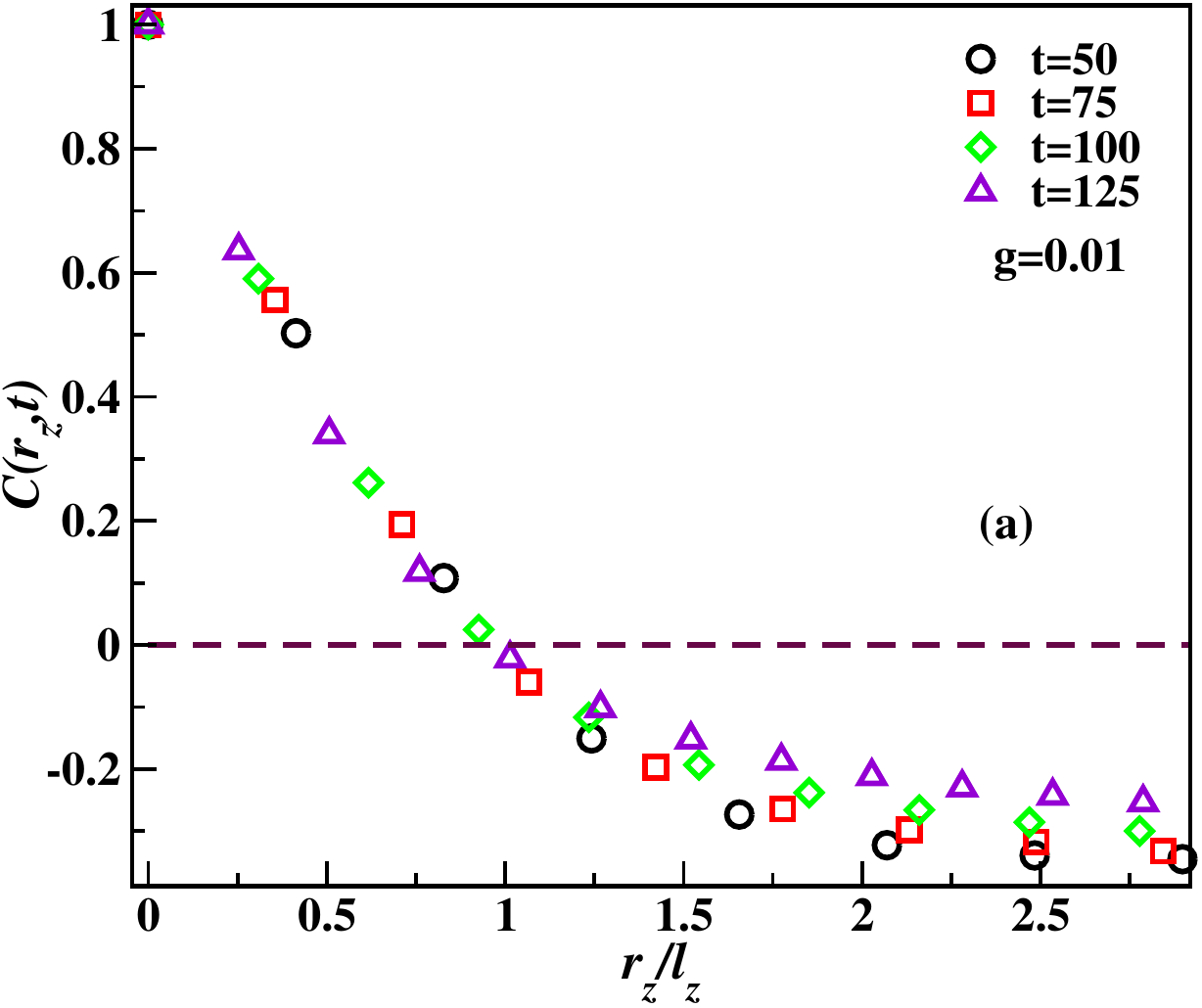}}\\
	{\includegraphics[width=0.45\textwidth]{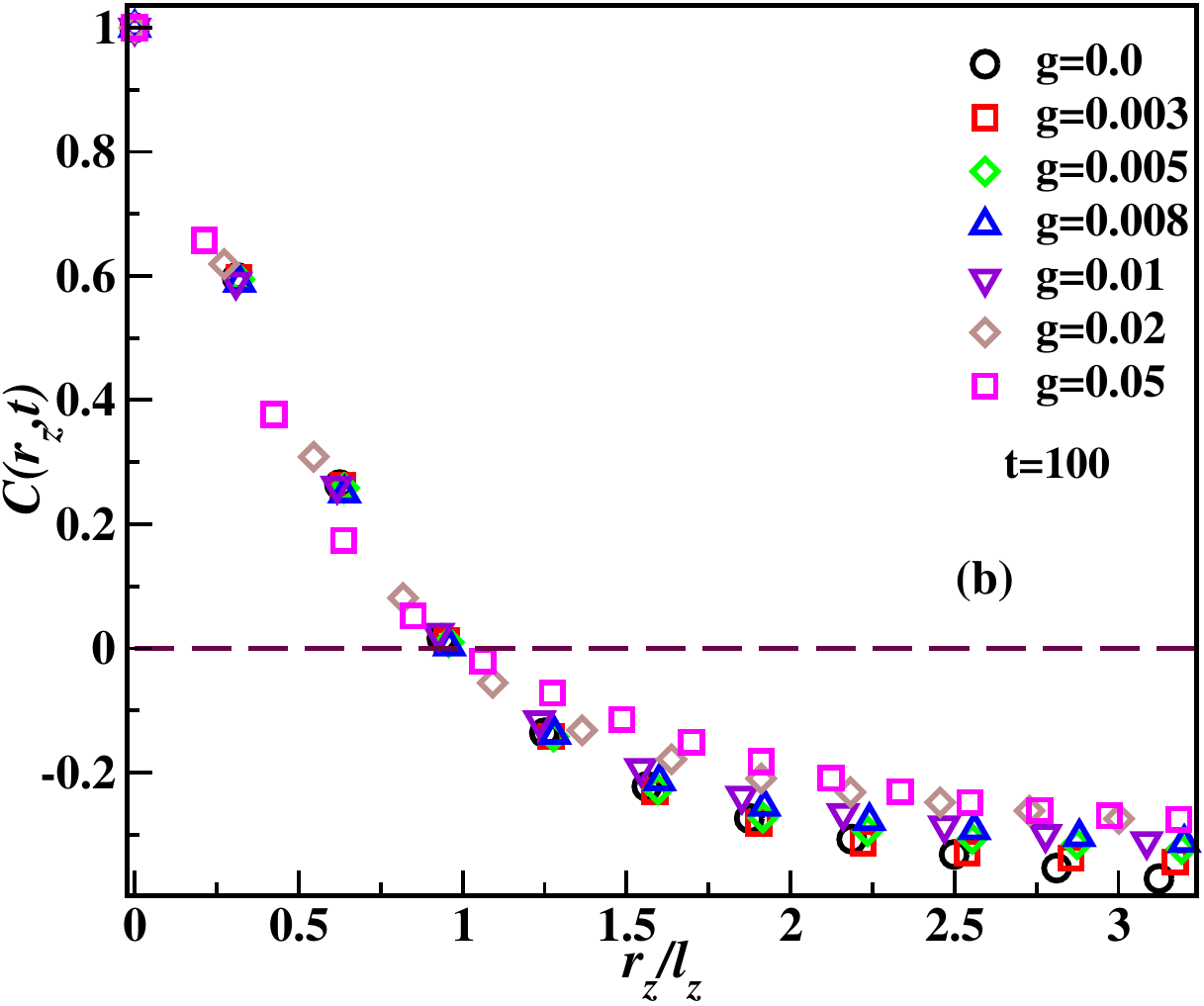}}
	\caption{The scaling plots of $C(r_z,t)$ vs $r_z/\ell_z(t)$ for $g=0.01$ at different times are shown in (a). In (b) the same scaling plots are shown for different $g$ values at a fixed time $t=100$.}
	\label{fig11:solid_correlation}
\end{figure}

The meaningful quantification of growth dynamics and extraction of the mentioned length scale depend on the satisfaction of the scaling property by the correlation function $C(r,t)$. In Fig.~\ref{fig11:solid_correlation}a, we present a scaling plot of $C(r,t)$ at a selected value of $g=0.01$ for various times. The result clearly illustrates the scaling behavior of the correlation functions in the region $r_z/\ell_z<1.0$, indicating that the domain growth is inherently self-similar. The lack of scaling in $C(r,t)$ at larger distance is attributed to the asymmetry in the composition concerning negative and positive values of the order parameter. This asymmetry increases with time as the domains compactify in the presence of gravity. We performed the same analysis for different $g$ values (not depicted here), and consistently observed a similar behavior, reinforcing the validity of the length scale analysis. 

Furthermore, to assess the validity of Porod law and superuniversality in our system, we plot in Fig.~\ref{fig11:solid_correlation}b the $C(r,t)$ for all the chosen g values at a fixed time. For zero gravity case, the correlation function exhibits a linear behavior at small distance as $C(z,t) \simeq 1 - az + \cdots$ due to the scattering from sharp interfaces, popularly known as Porod law. Therefore, the scaling of $C(r,t)$ displayed in Fig.~\ref{fig11:solid_correlation}b showing a linear trend for $r_z/\ell_z<1.0$ affirms the validity of Porod law in the presence of gravity. Additionally, the overlap in the correlation functions for different g values confirms the existence of superuniversality in the system. Note that, with increasing field strength, the asymmetry in the order parameter increases, resulting in the lack of scaling at large distance.  
\begin{figure}[ht!]
	\centering
	{\includegraphics[width=0.45\textwidth]{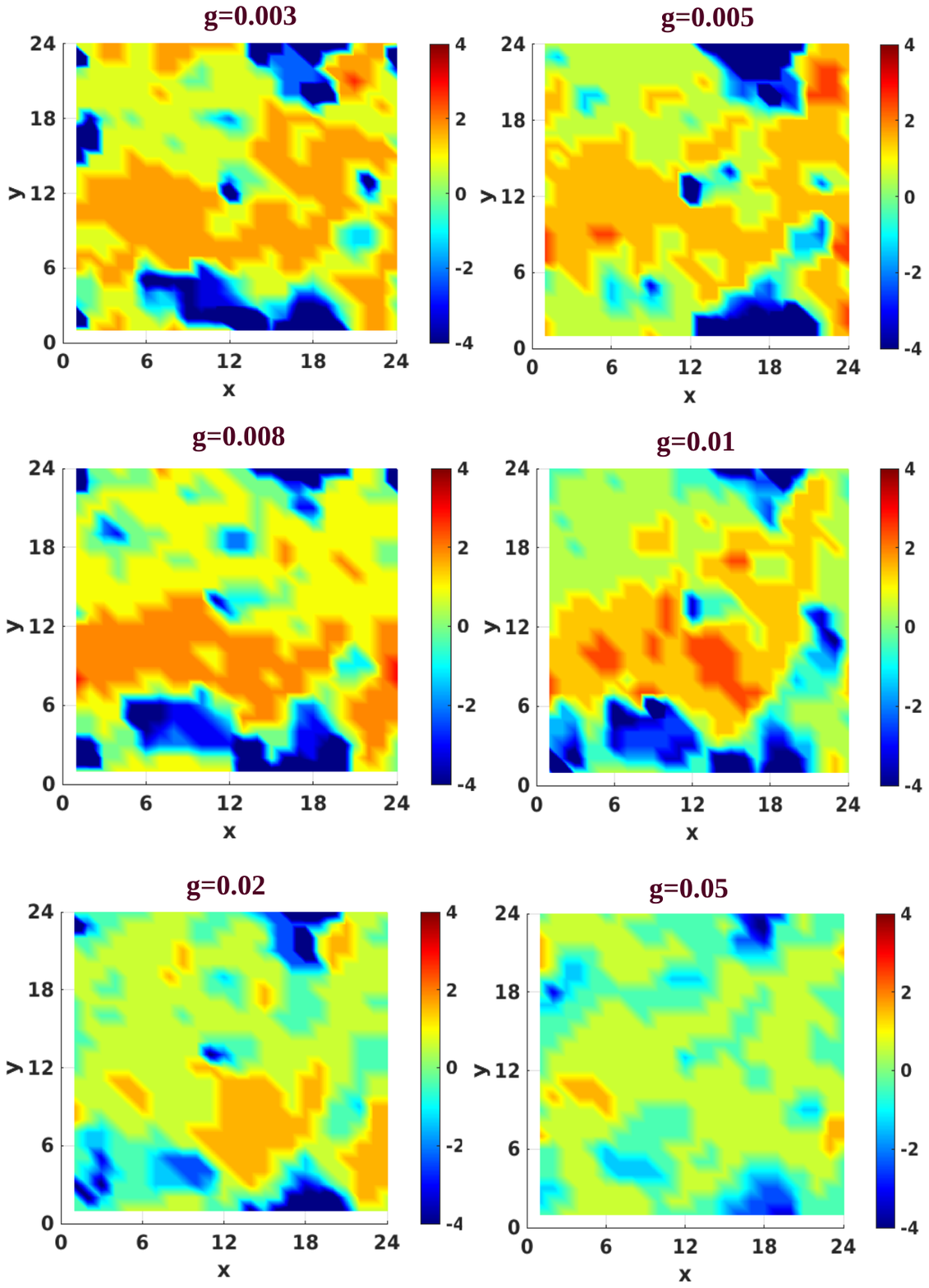}}
	\caption{Color maps of the height fields for the vapor-solid interface with respect to its mean for different $g$ values. The colors represent the local interface height in relation to its mean.}
	\label{fig12:surface roughness}
\end{figure}

Finally, we explore the influence of gravity on the vapor-solid interface at the final stage of the system that has undergone phase separation over a long period. In contrast to vapor-liquid phase separation, where the two phases exhibit a smooth interface, in the current scenario, with the accumulation of the solid phase at the bottom, the interface takes on a rough texture. The gravitational force is anticipated to play a role in modulating the structure of this interface. The resulting height variation is shown in Fig.~\ref{fig12:surface roughness} using a color map. Here we show the top view of the typical interfaces for all the $g$ values under consideration. For the low gravity case a pronounced roughness across the interface is visible. As the gravitational strength is increased the color map becomes homogeneous. This is a clear indication that the height fluctuation  decrease with increasing gravity strength, symbolizing a reduction in the surface roughness.

\begin{figure}[ht!]
	\centering
	{\includegraphics[width=0.45\textwidth]{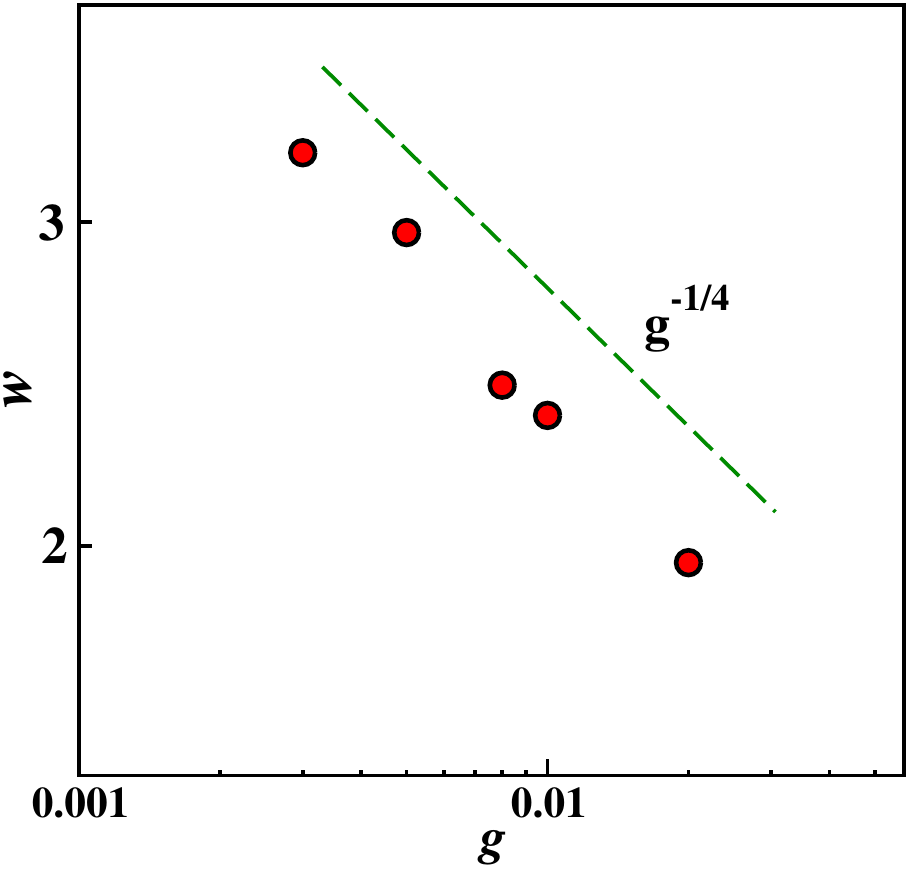}}
	\caption{The standard deviation of the vapor-solid interface roughness as a function of gravity. The dotted line is the guideline for the slope.}
	\label{fig13:std_solid}
\end{figure}

To quantify the interface roughness, we compute the standard deviation of the surface height distribution given by
\begin{equation}
	w = \sqrt{\langle h(x,y)^2 \rangle - \langle h(x,y) \rangle^2}
\end{equation}
The results are shown in Fig.~\ref{fig13:std_solid} with gravity. The interface roughness diminishes with a rise in gravitational strength, aligning with a power-law relationship of $g^{-1/4}$. Therefore, the power-law behavior suggests that the roughness of the interface approaches zero as gravitational strength becomes exceedingly large.

\section{Summary and Discussion}
In summary, we have employed molecular dynamics simulations to study the effect of gravity on the kinetics of vapor-liquid as well as vapor-solid phase separation of a one component Lennard Jones fluid. The phase separation was initiated by the thermal quench of the initially prepared high temperature homogeneous fluid inside the coexistence region. The hydrodynamics of the fluid was preserved by using Nose-Hoover thermostat. The density of the liquid being close to the critical value, we observe a bicontinuous domain structure of all the phases. To characterize the domain morphology and study the domain growth we resort to the two-point equal time order parameter correlation function. In the absence of gravity, our systems clearly exhibit the domain growth law consistent with the existing literature. However, when the systems are subjected to gravity, they become anisotropic, and a faster domain growth is observed along the direction of the field. The growth mechanics resemble the sedimentation process. After an initial period, under the influence of gravity, the scaling law breaks down and a new length scale emerges which strongly depends on the field strength.  We examined the validity of the Porod law and superuniversality in terms of the correlation function and the static structure factor. Our results demonstrated that the Porod's law and superuniversality remained vindicated in the presence of gravity. Finally, we studied the equilibrium state of the phase separated systems at long time. At this point, the complete understanding of the scaling exponents with gravitational field is missing and it will certainly be worthwhile to investigate. Our model can easily be extended to study the effect of gravity on the coarsening dynamic in many other segregating systems e.g. liquid-liquid phase separation, and it will be worth exploring these. 

\subsection*{Acknowledgements} 
B. Sen Gupta acknowledges Science and Engineering Research Board (SERB), Department of Science and Technology (DST), Government of India (no. CRG/2022/009343) for financial support. D. Davis acknowledges VIT for doctoral fellowship. Discussions with Parameshwaran A. are gratefully acknowledged.


\end{document}